\documentclass[a4paper,11pt]{article}
\usepackage{amssymb,graphicx,amsmath}

\makeatletter 
\renewcommand\section{\@startsection{section}{1}{\z@}              {-3.25ex\@plus -1ex \@minus -.2ex}                                    {1.5ex \@plus .2ex}                                    {\reset@font\Large\bfseries}}
\renewcommand\subsection{\@startsection{subsection}{2}{\z@}                                    {3.25ex \@plus 1ex \@minus.2ex}                                    {-1em}                                    {\reset@font\large\bfseries}}
\@addtoreset{equation}{section} \makeatother
\renewcommand{\theequation}{\thesection.\arabic{equation}}
\input{tcilatex}

\newcommand{\bc}{\begin{center}}
\newcommand{\ec}{\end{center}}
\def\ba#1{\begin{array}{#1}\displaystyle}
\newcommand{\ea}{\end{array}}

\newcommand{\beq}{\begin{equation}}
\newcommand{\eeq}{\end{equation}}
\newcommand{\beqa}{\begin{eqnarray}}
\newcommand{\eeqa}{\end{eqnarray}}
\newcommand{\no}{\nonumber}
\newcommand{\n}{\nonumber\\}
\newcommand{\bi}{\begin{itemize}}
\newcommand{\ei}{\end{itemize}}
\def\mato#1{\left(\ba{#1}} 
\def\matf{\ea\right)}

\def\lt#1{\left#1}
\def\rt#1{\right#1}
\def\t#1{\tilde{#1}}
\def\h#1{\hat{#1}}
\def\b#1{\bar{#1}}
\def\frc#1#2{\frac{#1}{#2}}

\newcommand{\p}{\partial}

\newcommand{\vac}{{\rm vac}}
\newcommand{\bra}{\langle}
\newcommand{\ket}{\rangle}

\newcommand{\R}{{\mathbb{R}}}

\newcommand{\Or}{{\cal O}}

\newcommand{\ep}{\epsilon}
\newcommand{\varep}{\varepsilon}

\newcommand{\Tr}{{\rm Tr}}

\newcommand{\tw}{{\cal T}}
\newcommand{\orb}{{\cal M}}
\newcommand{\sym}{\sigma}

\newcommand{\la}{\lambda}
\newcommand{\rx}{{\rm x}}
\newcommand{\ry}{{\rm y}}

\begin{document}

\setcounter{page}{0} \topmargin0pt \oddsidemargin0mm \renewcommand{%
\thefootnote}{\fnsymbol{footnote}} \newpage \setcounter{page}{0}
\begin{titlepage}

\vspace{0.2cm}
\begin{center}
{\Large {\bf Form factors of branch-point twist fields in quantum
integrable models and entanglement entropy }}

\vspace{0.8cm} {\large  \text{J.L.~Cardy$^{\circ}$,
O.A.~Castro-Alvaredo$^{\bullet}$ and B.~Doyon$^{\star}$}}

\vspace{0.2cm} {$^{\circ \, \star}$  Rudolf Peierls Centre for
Theoretical Physics, Oxford University,\\ 1 Keble Road, Oxford OX1
3NP, UK }\\ \vspace{0.2cm}
{$^{\circ}$ All Souls College, Oxford
OX1 4AL, UK}\\ \vspace{0.2cm}
{$^{\bullet}$  Centre for Mathematical Science, City University London, \\
Northampton Square, London EC1V 0HB, UK}
\end{center}
\vspace{1cm}

In this paper we compute the leading correction to the bipartite
entanglement entropy at large sub-system size, in integrable
quantum field theories with diagonal scattering matrices. We find
a remarkably universal result, depending only on the particle
spectrum of the theory and not on the details of the scattering
matrix. We employ the ``replica trick" whereby the entropy is
obtained as the derivative with respect to $n$ of the trace of the
$n^{{\rm th}}$ power of the reduced density matrix of the
sub-system, evaluated at $n=1$. The main novelty of our work is
the introduction of a particular type of twist fields in quantum
field theory that are naturally related to branch points in an
$n$-sheeted Riemann surface. Their two-point function directly
gives the scaling limit of the trace of the $n^{{\rm th}}$ power
of the reduced density matrix. Taking advantage of integrability,
we use the expansion of this two-point function in terms of form
factors of the twist fields, in order to evaluate it at large
distances in the two-particle approximation. Although this is a
well-known technique, the new geometry of the problem implies a
modification of the form factor equations satisfied by standard
local fields of integrable quantum field theory. We derive the new
form factor equations and provide solutions, which we specialize
both to the Ising and sinh-Gordon models.

\vfill{ \hspace*{-9mm}
\begin{tabular}{l}
\rule{6 cm}{0.05 mm}\\
$^\circ \text{j.cardy1@physics.ox.ac.uk}$\\
$^\bullet \text{o.castro-alvaredo@city.ac.uk}$\\
$^\star \text{b.doyon1@physics.ox.ac.uk}$\\
\end{tabular}}

\renewcommand{\thefootnote}{\arabic{footnote}}
\setcounter{footnote}{0}

\end{titlepage}
\newpage
\section{Introduction}

Quantum field theory (QFT) has proven to be one of the most
successful theories of the physical world. Its main objects are
correlation functions of local fields: they describe quantum
correlations between separated local observables and provide all
physical information that can be extracted from a model of QFT. In
general, correlation functions can be computed only
perturbatively, giving expressions that apply only in the
large-energy region. In contrast, some two-dimensional QFTs (one
space and one time dimension) have the property of integrability,
which means that correlation functions are accessible
non-perturbatively. In particular, their low-energy or
large-distance behavior is accessible in exact form in many cases,
from exact expressions for so-called form factors of local fields
\cite{KW,smirnovbook}. This fact has triggered an enormous amount
of work in computing form factors in a multitude of models of
integrable QFT (IQFT), some of which has found applications to
low-dimensional condensed matter systems \cite{Essler:2004ht}.

The main characteristic of a quantum system, as opposed to a
classical one, is the existence of entanglement: performing a
local measurement may instantaneously affect local measurements
far away. This property is essential to the field of quantum
computation and teleportation. At the theoretical level, there has
been considerable interest in formulating measures of quantum
entanglement \cite{bennet}-\cite{Verstraete} and applying them to
extended quantum systems with many degrees of freedom, such as
quantum spin chains \cite{Eisert}-\cite{Weston}. One of these
measures is entanglement entropy \cite{bennet}. Consider a quantum
system, with Hilbert space ${\cal H} = {\cal H}_A\otimes{\cal
H}_B$, in a pure state $|\psi\rangle$. The bipartite entanglement
entropy $S_A$ is the von Neumann entropy associated to the reduced
density matrix of the subsystem $A$, defined as
 \beq
    \rho_A = \Tr_{{\cal H}_{B}}(|\psi\ket\bra \psi|)\,,
\eeq
 \beq
    S_A = -\Tr_{{\cal H}_A} (\rho_A \log(\rho_A))~.
    \label{entropy}
\eeq

A recent application of QFT has been to the calculation of
entanglement entropy for the case of one-dimensional
systems \cite{Calabrese:2004eu, Calabrese:2005in}, extending
primarily on the work \cite{HolzheyLW94}, where ${\cal H}_A$ is
spanned by the degrees of freedom in some interval $A$ (or set of
intervals) of the real line, and $B$ is its complement. The
authors evaluated the bipartite entanglement entropy in quantum
systems at criticality, using techniques of conformal field theory
(CFT), as well as the leading large-distance limit in a massive
QFT. For example, when $|\psi\rangle$ is the ground state and $A$
is an interval of length $r$ in an infinite system, they found
\beq\label{001}
    S_A \sim \lt\{ \ba{ll} \frc{c}3 \log(r/\epsilon) & \epsilon\ll r\ll m^{-1} \\ \displaystyle
        -\frc{c}3 \log(\epsilon m) & r \gg m^{-1} \ea \rt.
\eeq where $m^{-1}$ is a correlation length of the QFT, $\epsilon$
is some short-distance cutoff, and $c$ is the central charge of
the CFT.

In this paper, we will develop a framework for the computation of
entanglement entropy in massive IQFT using factorized scattering
techniques. Our main result is the form of the first sub-leading
corrections to the entanglement entropy at large distances.
Remarkably, we show that the leading $r$-dependent correction is
independent of the precise details of the $S$-matrix, being
entirely determined by the spectrum of masses of the IQFT whenever
the scattering between particles does not involve backscattering.
That is, \beq
    S_A =-\frac{c}{3}\log(\epsilon m_1) + U -  \frc18 \sum_{\alpha=1}^\ell K_0(2rm_\alpha) + O\lt(e^{-3rm_1}\rt)
\eeq where $\alpha$ labels the particle types, $m_\alpha$ are the
associated masses (with $m_1$ the smallest one), $U$ is a
model-dependent constant, and $K_0(z)$ is the modified Bessel
function. For free theories, this result was previously obtained
by a different approach \cite{casini1, casini2}. The constant $U$
depends on the definition of $\ep$, but it can be fixed by
requiring that no constant correction terms occur, for instance,
in the upper expression in (\ref{001}). With such a definition, we
evaluated $U$ in the quantum Ising model, with the result: \beq
    U_{{\rm Ising}} = \frc16 \log 2 - \int_0^\infty \frc{dt}{2t}
    \lt( \frc{t\cosh t}{\sinh^3 t} - \frc1{\sinh^2t} - \frc{e^{-2t}}3\rt) = -0.131984...\label{ehh}
\eeq which, as we will explain in section \ref{sectentang}, is in
very good agreement with existing numerical results
\cite{Latorre2}. This constant could in principle also be
recovered by numerically integrating the Painlev\'e V equation
according to the results obtained in \cite{casini1}.

The main novelty of our approach is the introduction of a certain
type of twist field $\tw$ in IQFT, whose two-point function is
directly related to the entanglement entropy. The initial idea
\cite{Calabrese:2004eu} is to evaluate the entanglement entropy by
a ``replica trick" from the partition function on a multi-sheeted
Riemann surface. The field we introduce naturally arises as a
local field in an $n$-copy version of a given model of IQFT, and
implements branch points so that its correlation functions are
partition functions on multi-sheeted Riemann surfaces. The
large-distance corrections to the bipartite entanglement entropy
is derived from a large-distance expansion of the two-point
function of $\tw$, obtained by evaluating its form factors. Due to
the new geometry of the problem,  the form factor equations differ
from their usual form derived thirty years ago \cite{KW,
smirnovbook}. The operation of evaluating the entanglement entropy
from the result of the form factor expansion involves a subtle
analytic continuation in $n$. We provide a general derivation of
the form factor equations and of the entanglement entropy in the
context of diagonal factorised scattering theory, and we have
studied the Ising and sinh-Gordon cases in detail.

The paper is organised as follows: in section \ref{gen}, we
discuss partition functions  on multi-sheeted spaces and their
relation to entanglement entropy in general terms. In section
\ref{sectff}, we develop the form factor program for twist fields,
and provide a general formula for the two-particle form factors of
theories with a single particle spectrum and no bound states,
which we specialise to both the Ising and the sinh-Gordon cases.
In section \ref{identifying} we check our previous formulae for
consistency by computing the ultraviolet conformal dimension of
the twist field in the two-particle approximation both for the
Ising and sinh-Gordon model. In section \ref{sectentang} we
compute the two-point function of the twist field in the
two-particle approximation and derive from that our general
formula for the bipartite entanglement entropy for theories with a
single particle spectrum and no bound states. In section
\ref{manypar} we extend the previous results to diagonal theories
with many particles and bound states. In section \ref{conclusion}
we present our conclusions and point out some open problems.
Finally we provide four appendices: in appendix \ref{appshG} we give an
alternative derivation of the form factors of the twist field in
the $n$-copy sinh-Gordon theory using the method of angular
quantization; in appendix \ref{appising} we compute the vacuum expectation
value of the twist field in the $n$-copy Ising model and derive
the value (\ref{ehh}) from it; in appendix \ref{appf} we present the
details of the analytic continuation necessary for the computation
of the entropy and in appendix \ref{finf} we provide the $n
\rightarrow \infty$ limit of the two-point function of the twist
field in the sinh-Gordon model in the two-particle and
saddle-point approximation.

\section{Partition functions on multi-sheeted spaces and entanglement entropy}\label{gen}

\subsection{Partition functions in QFT on multi-sheeted spaces}\indent \\

\begin{figure}
\bc
\includegraphics[width=8cm,height=5cm]{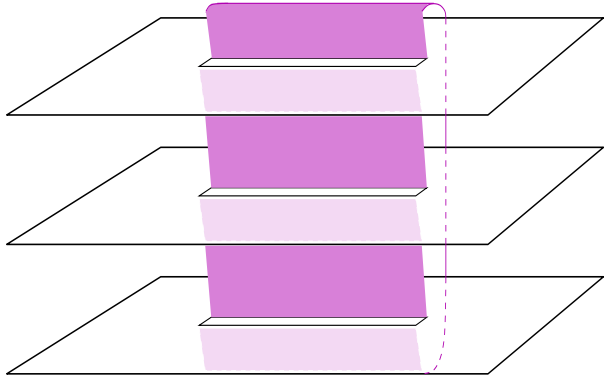}
\ec \caption{[Color online] A representation of the Riemann
surface $\orb_{3,a_1,a_2}$.} \label{fig-sheets}
\end{figure}

The partition function of a model of two-dimensional QFT with
local lagrangian density ${\cal L}[\varphi](\rx,\ry)$ on a
(euclidean-signature) Riemann surface ${\cal R}$ is formally
obtained by the path integral \beq\label{partfunct}
    Z[{\cal L},{\cal R}] = \int [d\varphi]_{{\cal R}} \exp\lt[-\int_{{\cal R}} d\rx d\ry\,{\cal L}[\varphi](\rx,\ry)\rt]
\eeq where $[d\varphi]_{{\cal R}}$ is an infinite measure on the
set of configurations of some field $\varphi$ living on the
Riemann surface ${\cal R}$ and on which the lagrangian density
depends in a local way. Consider Riemann surfaces with curvature
zero everywhere except at a finite number of points. Since the
lagrangian density does not depend explicitly on the Riemann
surface as a consequence of its locality, it is expected that this
partition function can be expressed as an object calculated from
a model on $\R^2$, where the structure of the Riemann surface is
implemented through appropriate boundary conditions around the
points with non-zero curvature. Consider for instance the simple
Riemann surface $\orb_{n,a_1,a_2}$ composed of $n$ sheets
sequencially joined to each other on the segment
$x\in[a_1,a_2],\,y=0$ (see Fig. \ref{fig-sheets} representing the
case $n=3$). We would expect that the associated
partition function involves certain ``fields''\footnote{Here, the
term ``field'' is taken in its most general QFT sense: it is an
object of which correlation functions -- multi-linear maps -- can be evaluated, and which
depends on a position in space -- parameters $\rx,\ry$ that
transform like coordinates under translation symmetries.} at
$(\rx,\ry)=(a_1,0)$ and $(\rx,\ry)=(a_2,0)$.

The expression (\ref{partfunct}) for the partition function
essentially defines these fields (that is, it gives their
correlation functions, up to a normalisation independent of their
positions). But in the model on $\R^2$, this definition makes them
non-local. Locality of a field (used here in its most fundamental
sense) means that as an observable in the quantum theory, it is
quantum mechanically independent of the energy density at
space-like distances. In the associated euclidean field theory,
this means that correlation functions involving this field and the
energy density are, as functions of the position of the energy
density, defined on $\R^2$ (and smooth except at the positions of
the fields). The energy density is simply obtained from the
lagrangian density, hence it is clear that fields defined by
(\ref{partfunct}) in the model on $\R^2$ are not local. Locality
is at the basis of most of the results in integrable QFT, so it is
important to recover it.

The idea is simply to consider a larger model: a model formed by
$n$ independent copies of the original model, where $n$ is the
number of Riemann sheets necessary to describe the Riemann surface
by coordinates on $\R^2$. Let us take again the simple example of
$\orb_{n,a_1,a_2}$. We re-write (\ref{partfunct}) as
\beq\label{partfunctmulti}
    Z[{\cal L},\orb_{n,a_1,a_2}] =
        \int_{{\cal C}(a_1;a_2)} [d\varphi_1 \cdots d\varphi_n]_{\R^2} \exp\lt[-\int_{\R^2} d\rx d\ry\,({\cal L}[\varphi_1](\rx,\ry)+\ldots+{\cal L}[\varphi_n](\rx,\ry))\rt]
\eeq where ${\cal C}(a_1,a_2)$ are {\em conditions} on the fields
$\varphi_1,\ldots,\varphi_n$ restricting the path integral: \beq
    {\cal C}(a_1;a_2) \quad:\quad \varphi_i(\rx,0^+) = \varphi_{i+1}(\rx,0^-)~,\quad \rx\in[a_1,a_2],\; i=1,\ldots,n
\eeq where we identify $n+i\equiv i$. The lagrangian density of the
multi-copy model is
\[
    {\cal L}^{(n)}[\varphi_1,\ldots,\varphi_n](\rx,\ry) = {\cal L}[\varphi_1](\rx,\ry)+\ldots+{\cal L}[\varphi_n](\rx,\ry)
\]
so that the energy density in that model is the sum of the energy
densities of the $n$ individual copies. Hence the expression
(\ref{partfunctmulti}) does indeed define local fields at
$(a_1,0)$ and $(a_2,0)$ in the multi-copy model, since this sum is
the same on both sides of the segment $\rx\in[a_1,a_2],\,\ry=0$
according to the conditions ${\cal C}(a_1,a_2)$.

The local fields defined in (\ref{partfunctmulti}) are examples of ``twist fields''. Twist fields exist in a QFT model whenever there is a global internal symmetry $\sym$ (a symmetry that acts the same way everywhere in space, and that does not change the positions of fields): $\int_{\R^2}d\rx d\ry\, {\cal L}[\sym\varphi](\rx,\ry) = \int_{\R^2}d\rx d\ry\, {\cal L}[\varphi](\rx,\ry)$. Their correlation functions can be formally defined through the path integral:
\beq
    \bra \tw_\sym(a,b) \cdots \ket_{{\cal L},\R^2} \propto
        \int_{{\cal C}_\sym(0,0)} [d\varphi]_{\R^2} \exp\lt[-\int_{\R^2} d\rx d\ry\,{\cal L}[\varphi](\rx,\ry)\rt] \cdots
\eeq where $\cdots$ represent insertions of other local fields at
different positions and the path integral conditions are \beq
    {\cal C}_\sym(a,b) \quad:\quad \varphi(\rx,b^+) = \sym\varphi(\rx,b^-)~,\quad \rx\in[a,\infty)~.
\eeq
The proportionality constant is an infinite constant that is independent of the position $(a,b)$ and of those of the other local fields inserted. The fact that $\sym$ is a symmetry ensures that $\tw_\sym$ is local. Also, it insures that the result is in fact independent of the shape of the cut in the conditions ${\cal C}_\sym$, up to symmetry transformations of the other local fields inserted. A consequence of this definition is that correlation functions $\bra \tw_\sym(a,b) \Or(\rx,\ry) \cdots\ket_{{\cal L},\R^2}$ with some local fields $\Or(\rx,\ry)$ are defined, as functions of $\rx,\ry$ (smooth except at positions of other local fields), on a multi-sheeted covering of $\R^2$ with a branch point at $(a,b)$, whenever $\sym\Or \neq\Or$. They have the property that a clockwise turn around $(a,b)$ is equivalent to the replacement $\Or\mapsto \sym\Or$ in the correlation function. If $\sym\Or \neq\Or$, then $\Or$ is said to be ``semi-local'' with respect to $\tw_\sym$. This property, along with the condition that $\tw_\sym$ has the lowest scaling dimension and be invariant under all symmetries of the model that commute with $\sym$ (that is, that it be a primary field in the language of conformal field theory), is expected to uniquely fix the field $\tw_\sym$, and constitute a more fundamental definition than the path integral above, as it does not require the existence of a lagrangian density. We will take this point of view in the following, but we will continue to denote a model of QFT by ${\cal L}$ and its $n$-copy tensor product by ${\cal L}^{(n)}$.

\begin{figure}
\bc
\includegraphics[width=6cm,height=3cm]{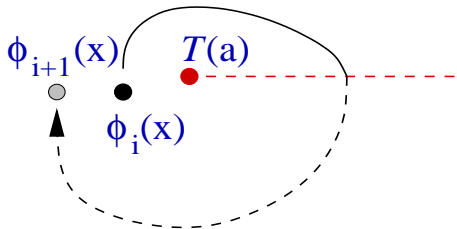}
\ec \caption{[Color online] The effect of $\tw$ on other local
fields.} \label{fig-Cx}
\end{figure}

In the model with lagrangian ${\cal L}^{(n)}$, there is a symmetry under exchange of the copies. The twist fields defined by (\ref{partfunctmulti}), which we call {\em branch-point twist fields}, are twist fields associated to the two opposite cyclic permutation symmetries $i\mapsto i+1$ and $i+1\mapsto i$  ($i=1,\ldots,n,\;n+1\equiv 1$). We will denote them simply by $\tw$ and $\t\tw$, respectively:
\beqa &&
    \tw=\tw_\sym~,\quad \sym\;:\; i\mapsto i+1 \ {\rm mod} \,n \n &&
    \t\tw=\tw_{\sym^{-1}}~,\quad \sym^{-1}\;:\; i+1\mapsto i \ {\rm mod} \,n\no
\eeqa
(see Fig. \ref{fig-Cx} for the case $\tw$).
More precisely, we have
\beq\label{pftwopt}
    Z[{\cal L},\orb_{n,a_1,a_2}] \propto \bra \tw(a_1,0) \t\tw(a_2,0) \ket_{{\cal L}^{(n)},\R^2}~.
\eeq
This can be seen by observing that for $\rx\in[a_1,a_2]$, consecutive copies are connected through $\ry=0$ due to the presence of $\tw(a_1,0)$, whereas for $\rx>a_2$, copies are connected to themselves through $\ry=0$ because the conditions arising from the definition of $\tw(a_1,0)$ and $\t\tw(a_2,0)$ cancel each other.

More generally, the identification holds for correlation functions in the model ${\cal L}$ on $\orb_{n,a_1,a_2}$, this time with an equality sign:
\beq\label{br-tw}
    \bra \Or(\rx,\ry; \mbox{sheet $i$}) \cdots \ket_{{\cal L},\orb_{n,a_1,a_2}} =
    \frc{\bra \tw(a_1,0) \t\tw(a_2,0) \Or_i(\rx,\ry) \cdots \ket_{{\cal L}^{(n)},\R^2}}{\bra \tw(a_1,0) \t\tw(a_2,0) \ket_{{\cal L}^{(n)},\R^2}}
\eeq
where $\Or_i$ is the field in the model ${\cal L}^{(n)}$ coming from the $i^{\rm th}$ copy of ${\cal L}$.

The generalisation to Riemann surfaces with more branch points is straightforward, but will not be needed here.

The conformal dimension of branch-point twist fields was calculated\footnote{In fact, in the paper \cite{Calabrese:2004eu} branch-point twist fields in the multi-copy model were not introduced explicitly, as they are not essential for the evaluation of partition functions in CFT. Only the non-local fields discussed above were alluded to, but the method to evaluate the scaling dimension is the same.} in \cite{Calabrese:2004eu}. Consider the model ${\cal L}$ to be a conformal field theory (CFT). Then also ${\cal L}^{(n)}$ is a CFT. There are $n$ fields $T_j(z)$ in ${\cal L}^{(n)}$ that correspond to the stress-energy tensors of the $n$ copies of ${\cal L}$, and in particular the sum $T^{(n)}(z) = \sum_{j=1}^n T_j(z)$ is the stress-energy tensor of ${\cal L}^{(n)}$. The central charge of ${\cal L}^{(n)}$ is $nc$, if $c$ is that of ${\cal L}$.

Consider the stress-energy tensor $T(w)$ in ${\cal L}$. We can evaluate the one-point function $\bra T(w)\ket_{{\cal L},\orb_{n,a_1,a_2}}$ by making a conformal transformation from $z$ in $\R^2$ to $w$ in $\orb_{n,a_1,a_2}$ (here $z$ and $w$ are complex coordinates, with for instance $z=\rx+i\ry$) given by
\[
    z = \lt(\frc{w-a_1}{w-a_2}\rt)^{\frc1n}~.
\]
We have
\[
    \bra T(w)\ket_{{\cal L},\orb_{n,a_1,a_2}} = \lt(\frc{\p z}{\p w} \rt)^2 \bra T(z)\ket_{{\cal L},\R^2} + \frc{c}{12} \{z,w\}
\]
where the Schwarzian derivative is
\[
    \{z,w\} = \frc{z''' z' - (3/2) (z'')^2}{(z')^2}~.
\]
Using $\bra T(z)\ket_{{\cal L},\R^2}=0$, we obtain
\[
    \bra T(w)\ket_{{\cal L},\orb_{n,a_1,a_2}} = \frc{c(n^2-1)}{24 n^2} \frc{(a_1-a_2)^2}{(w-a_1)^2(w-a_2)^2} ~.
\]
Since, by (\ref{br-tw}), this is equal to $\bra \tw(a_1,0) \t\tw(a_2,0) T_j(w)\ket_{{\cal L}^{(n)},\R^2}/\bra \tw(a_1,0) \t\tw(a_2,0)\ket_{{\cal L}^{(n)},\R^2}$ for all $j$, we can evaluate the correlation function involving the stress-energy tensor of ${\cal L}^{(n)}$ by multiplying by $n$:
\[
    \frc{\bra \tw(a_1,0) \t\tw(a_2,0) T^{(n)}(w)\ket_{{\cal L}^{(n)},\R^2}}{\bra \tw(a_1,0) \tw(a_2,0)\ket_{{\cal L}^{(n)},\R^2}} =
    \frc{c(n^2-1)}{24 n} \frc{(a_1-a_2)^2}{(w-a_1)^2(w-a_2)^2}~.
\]
From the usual CFT formula for insertion of a stress-energy tensor
\beqa &&
    \bra \tw(a_1,0) \t\tw(a_2,0) T^{(n)}(w)\ket_{{\cal L}^{(n)},\R^2} = \n && \qquad    \lt(\frc1{w-a_1} \frc{\p}{\p a_1} + \frc{h_1}{(w-a_1)^2} + \frc1{w-a_2} \frc{\p}{\p a_2} + \frc{h_2}{(w-a_2)^2}\rt)
    \bra \tw(a_1,0) \t\tw(a_2,0)\ket_{{\cal L}^{(n)},\R^2} \no
\eeqa
we identify the scaling dimension of the primary fields $\tw$ and $\t\tw$ (they have the same scaling dimension) using $\bra \tw(a_1,0) \t\tw(a_2,0)\ket_{{\cal L}^{(n)},\R^2} = |a_1-a_2|^{-2d_n}$:
\beq\label{scdim}
    d_n = \frc{c}{12} \lt(n-\frc1n\rt)~.
\eeq
It may happen that many fields with the main property of branch-point twist fields exist, with different dimensions. However, the dimension (\ref{scdim}) should be the lowest possible dimension. Hence, a field with the main properties of branch-point twist field, with this dimension, and invariant under all symmetries of the theory should be unique.

\subsection{Entanglement entropy}\indent \\

Partition functions on Riemann surfaces with branch points can be
used in order to evaluate the entanglement entropy; this works
when $A$ consists of one and also of more than one interval, as was explained in
\cite{Calabrese:2004eu}. Consider a (finite or infinite)
one-dimensional quantum system with Hilbert space ${\cal H} =
{\cal H}_A\otimes{\cal H}_B$ where ${\cal H}_A$ is the space of
local degrees of freedom in some interval (or set of intervals)
$A$. Consider also the ground state $|\psi\ket$ of this quantum
system. The bipartite entanglement entropy $S_A$ is defined as
follows. We first define the induced density matrix $\rho_A$ as
\beq
    \rho_A = \Tr_{{\cal H}_{B}}(|\psi\ket\bra \psi|)
\eeq
and then we calculate the von Neumann entropy associated to this density matrix:
\beq
    S_A = -\Tr_{{\cal H}_A} (\rho_A \log(\rho_A))~.
\eeq This has the interpretation of counting the number of fully
entangled ``links'' between the regions $A$ and $\b{A}$ as encoded
into the ground state $|\psi\ket$. In agreement with this
interpretation, it has the symmetry property $S_A = S_{\b{A}}$,
and in a quantum model with local interaction, it is expected to
saturate to a finite value when both regions $A$ and $\b{A}$ are
much larger than the correlation length.

The main idea in order to evaluate the entanglement entropy in the
scaling limit of a quantum model (here, we will consider models on
infinite space only) is to use the ``replica trick''
\cite{Calabrese:2004eu}. That is, we evaluate \beq\label{thetrace}
    \Tr_{{\cal H}_A} \rho_A^n
\eeq
then take the limit $n\to 1$ of the derivative with respect to $n$, using the identity $\rho_A \log\rho_A = \lim_{n\to 1} \frc{\p}{\p n}\rho_A^n$:
\beq\label{SA}
    S_A = -\lim_{n\to 1} \frc{d}{d n} \Tr_{{\cal H}_A} \rho_A^n~.
\eeq This formula requires that we evaluate the trace with real
positive $n$, but the trick is to evaluate $\Tr_{{\cal H}_A}
\rho_A^n$ with positive integer $n$, then to take the appropriate
``analytic continuation'' (it is unique if we assume a certain
asymptotic behaviour as $n\to\infty$ -- we will discuss this in
section \ref{sectentang}). Considering positive integer $n$ is
useful, because in the scaling limit, denoting the QFT model
associated to the region near the critical point by ${\cal L}$, we
have, taking $A$ to consist of only one interval,
\beq
    \Tr_{{\cal H}_A} \rho_A^n \to Z[{\cal L};\orb_{n,a_1,a_2}]
\eeq where $a_1$ and $a_2$ are the dimensionful end-points of the
region $A$ (the scaling limit is taken with the length of the
region $A$ in proportion to the correlation length, which is then sent to
infinity). As we saw above, this can be computed as a two-point
correlation function of local fields in ${\cal L}^{(n)}$ using
(\ref{pftwopt}). More precisely, with $m$ a mass scale of the QFT
and $\ep$ the some dimensionful distance of the order of the
site spacing, $m\ep$ being in
inverse proportion to the dimensionless correlation length, we
have \beq\label{rhon}
    \Tr_{{\cal H}_A} \rho_A^n \sim {\cal Z}_n\ep^{2d_n} \bra \tw(a_1,0) \t\tw(a_2,0) \ket_{{\cal L}^{(n)},\R^2}
\eeq with an $n$-dependent non-universal normalisation constant
${\cal Z}_n$ (with ${\cal Z}_1=1$), and where $d_n$ is the scaling
dimension (\ref{scdim}). For later convenience, $\ep$ is chosen
in such a way that $d {\cal Z}_n/dn =1$ at $n=1$. Note that the expression above
is dimensionless, since the operators $\tw$ and $\t\tw$ both have
dimension $d_n$ (in particular, their individual vacuum expectation value in the QFT
is proportional to $m^{d_n}$).

Similarly, for regions composed of many disconnected components,
$\Tr_{{\cal H}_A} \rho_A^n$ is identified with partition functions
on Riemann surfaces with many branch points, as explained in
\cite{Calabrese:2004eu}, but we will not consider this case here.

\section{The form factor program for branch-point twist fields} \label{sectff}

We now turn to the description of QFT on Minkowski space-time in
terms of its Hilbert space of asymptotic relativistic particles.
In the context of $1+1$-dimensional IQFT, form factors are defined as
tensor valued functions representing matrix elements of some local
operator $\mathcal{O}({x})$ located at the origin $x=0$ between a
multi-particle {\em{in}}-state and the vacuum:
\begin{equation}
F_{k}^{\mathcal{O}|\mu _{1}\ldots \mu _{k}}(\theta _{1},\ldots
,\theta _{k}):=\left\langle 0|\mathcal{O}(0)|\theta_1,\ldots,\theta_k\right\rangle_{\mu_1,\ldots,\mu_k}^{\text{in}} ~.\label{ff}
\end{equation}
Here $|0\rangle$ represents the vacuum state and
$|\theta_1,\ldots,\theta_k\rangle_{\mu_1,\ldots,\mu_k}^{\text{in}}$ the
physical ``in'' asymptotic states of massive QFT. They carry indices
$\mu_i$, which are quantum numbers characterizing the various
particle species, and depend on the real parameters $\theta_i$, which
are called rapidities. The form factors are defined for all rapidities by analytically continuing
from some ordering of the rapidities; a fixed ordering provides a complete basis of states.

The main characteristics of massive integrable models of QFT are
that the number of particles and their  momenta set are conserved
under scattering, and that the scattering matrix factorises into
products of two-particle scattering matrices, which are the
solutions of a set of consistency equations and analytic
properties. These consistency equations and analytic properties
are often strong enough to completely fix the scattering matrix in
integrable models. Similarly, the form factors are fixed by a set
of equations and analytic properties depending on the two-particle
scattering matrix (or $S$-matrix).

\begin{figure}
\bc
\includegraphics[width=6cm,height=5cm]{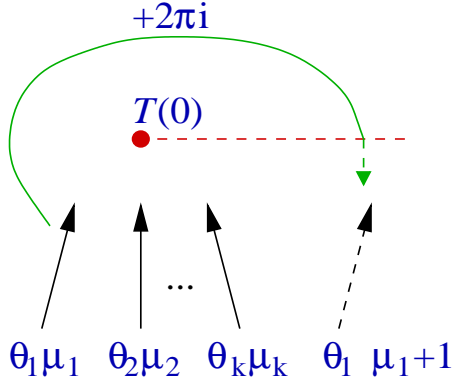}
\ec \caption{[Color online] A pictorial representation of the
effect of adding $2\pi i$ to rapidity $\theta_1$ in form factors
of the twist field $\tw$.} \label{fig-periodicity}
\end{figure}

\begin{figure}
\bc
\includegraphics[width=14cm,height=5cm]{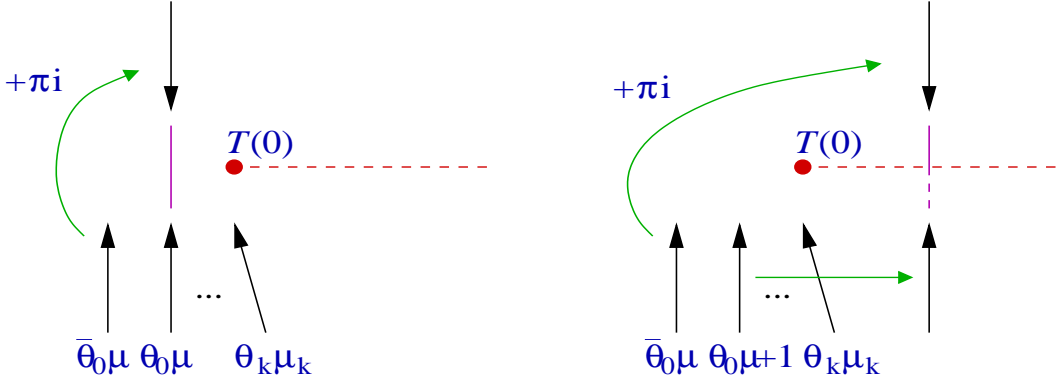}
\ec \caption{[Color online] The kinematic poles come from the
structure of the wave function far from the local fields, at
positive and negative infinity. Adding $i\pi$ to rapidity
$\theta_1$ puts the particle in the ``out" region. With a particle
in that region, there are delta-functions representing particles
in the ``in" region going through without interacting with the
local fields. Those occur from the $e^{ipx}$ form of the wave
function at positive and negative infinity. But if the
coefficients at both limits are different, $S_- e^{ipx}$ and
$S_+e^{ipx}$ with $S_-\neq S_+$, then there are also poles in
addition to these delta-functions. Only these poles are seen in
the analytic continuation $\theta_1\mapsto\theta_1+i\pi$.
Different coefficients come from the semi-locality of the twist
field and the non-free scattering matrix, as represented here.}
\label{fig-kinematic}
\end{figure}

In this section we want to show how the standard form factor
equations for $1+1$-dimensional IQFTs must be modified for the
branch-point twist fields. Let us consider an integrable model
consisting of $n$ copies of a known integrable theory possessing a
single particle spectrum and no bound states (such as the Ising
and sinh-Gordon models). We have therefore $n$ particles, which we
will denote by indices $1, \ldots, n$. The $S$-matrix between
particles $i$ and $j$ with rapidities $\theta_i$ and $\theta_j$
will be denoted by $S_{ij}(\theta_i-\theta_j)$ (that it depends on
the rapidity difference is a consequence of relativistic
invariance). Particles of different copies do not interact with
each other, so that the $S$-matrix of the model will be of the
form
\begin{eqnarray}\label{s}
   S_{ii}(\theta)&=& S(\theta) \qquad \forall \quad i=1,\ldots, n,\\
   S_{ij}(\theta)&=& 1, \qquad \forall \quad i,j=1, \ldots, n \quad \text{and} \quad i\neq j,
\end{eqnarray}
where  $S(\theta)$ is the $S$-matrix of the single-copy integrable
QFT. As explained above, as a consequence of the symmetry of the
model, a twist field $\mathcal{T}$ must exist such that if
$\Psi_{1}, \ldots, \Psi_{n}$ are the fields associated to the
fundamental particles of each copy of the original model, then the
equal time ($x^0=y^0$) exchange relations between $\mathcal{T}$
and $\Psi_{1}, \ldots, \Psi_{n}$ can be written in the following
form\footnote{Here we employ the standard notation in Minkowski
space-time: $x^{\nu}$ with $\nu=0,1$, with $x^{0}$ being the time
coordinate and $x^{1}$ being the position coordinate.}
\begin{eqnarray}
    \Psi_{i}(y)\mathcal{T}(x) &=& \mathcal{T}(x) \Psi_{i+1}(y) \qquad x^{1}> y^{1}, \n
    \Psi_{i}(y)\mathcal{T}(x) &=& \mathcal{T}(x) \Psi_{i}(y) \qquad x^{1}< y^{1}, \label{cr}
\end{eqnarray}
for $i=1,\ldots, n$ and where we identify the indices $n+i \equiv
i$. The relation with the previous section is obtained by
recalling that in going from the Hilbert space description to the
path integral description, the order of operators is translated
into time-ordering (or $\ry$-ordering in euclidean space), and
that left-most operators are later in time. It is well known that
such exchange relations play an important role in the derivation
of the consistency equations for the form factors of the operator
$\mathcal{T}$. Generalising the standard arguments to the exchange
relation (\ref{cr}), the form factor axioms are
\begin{eqnarray}
  F_{k}^{\mathcal{T}|\ldots \mu_i  \mu_{i+1} \ldots }(\ldots,\theta_i, \theta_{i+1}, \ldots ) &=&
  S_{\mu_i \mu_{i+1}}(\theta_{i\,i+1})
  F_{k}^{\mathcal{T}|\ldots \mu_{i+1}  \mu_{i} \ldots}(\ldots,\theta_{i+1}, \theta_i,  \ldots ), \n
 F_{k}^{\mathcal{T}|\mu_1 \mu_2 \ldots \mu_k}(\theta_1+2 \pi i, \ldots,
\theta_k) &=&
  F_{k}^{\mathcal{T}| \mu_2 \ldots \mu_n \hat{\mu}_1}(\theta_2, \ldots, \theta_{k},
  \theta_1)\n
 -i \text{Res}_{\substack{\bar{\theta}_{0}={\theta}_{0}}}
 F_{k+2}^{\mathcal{T}|\b{\mu} \mu  \mu_1 \ldots \mu_k}(\bar{\theta}_0+i\pi,{\theta}_{0}, \theta_1 \ldots, \theta_k)
  &=&
  F_{k}^{\mathcal{T}| \mu_1 \ldots \mu_k}(\theta_1, \ldots,\theta_k), \n
 -i \text{Res}_{\substack{\bar{\theta}_{0}={\theta}_{0}}}
 F_{k+2}^{\mathcal{T}|\b\mu \hat{\mu } \mu_1 \ldots \mu_k}(\bar{\theta}_0+i\pi,{\theta}_{0}, \theta_1 \ldots, \theta_k)
  &=&-\prod_{i=1}^{k} S_{{\mu}\mu_i}(\theta_{0i})
  F_{k}^{\mathcal{T}| \mu_1 \ldots \mu_k}(\theta_1, \ldots,\theta_k).\label{kre}
\end{eqnarray}
Here $\theta_{ij}=\theta_i-\theta_j$ and the first axiom is in
fact the same as for local fields. In the second equation, the
crossing or locality relation, we introduced the symbols
$\hat{\mu}_i=\mu_i+1$. As compared to the usual form factor
equations, it is altered by the nature of the exchange relation
and it now relates form factors associated to different particle
sets. Finally, the last two equations generalise the standard
kinematic residue equation to branch-point twist fields. Once
more, the exchange relations (\ref{cr}) are responsible for the
splitting into two equations. Here, for later convenience, we
wrote the equations in their general form valid also for
many-particle models, where $\b\mu$ represents the anti-particle
associated to $\mu$. In the present case, the integrable model we
started with has just one particle (so that $\mu$ labels the
copies) and therefore each particle is its own anti-particle.
Since we are dealing with theories without bound states, these are
in fact all the equations which one needs to solve. Pictorial
explanations of the second and of the last two equations are
given, respectively, in Figs. \ref{fig-periodicity} and
\ref{fig-kinematic}.

The other field $\tilde{\mathcal{T}}$ introduced in section \ref{gen} is a twist field with similar
properties as $\mathcal{T}$ but whose exchange relations with
the fundamental fields of the theory are given by
\begin{eqnarray}
    \Psi_{i}(y)\tilde{\mathcal{T}}(x) &=& \tilde{ \mathcal{T}}(x) \Psi_{i-1}(y) \qquad x^{1}> y^{1}, \n
    \Psi_{i}(y)\tilde{\mathcal{T}}(x) &=& \tilde{ \mathcal{T}}(x) \Psi_{i}(y) \qquad x^{1}< y^{1}, \label{cr2}
\end{eqnarray}
instead of (\ref{cr}).  This implies that on the Hilbert space, we have
\beq
    \t\tw = \tw^\dag~.
\eeq

In order to fully define the fields $\tw$ and $\t\tw$, we need to fix their normalisation, which does not follow from the form factor equations. We will adopt the usual CFT normalisation:
\beq\label{normalisationtw}
    \bra\tw(x)\t\tw(0)\ket \sim r^{-2d_n}\quad \mbox{as} \quad r\to0~.
\eeq
Here and below, $r$ denotes the space-like separation $\sqrt{(x^1)^2 - (x^0)^2}$, and the two-point function just depends on it thanks to relativistic invariance and spinless-ness of the fields involved.

\subsection{Two-particle form factors}\indent \\

\noindent As usual in this context, we define the minimal form
factors $F_{\text{min}}^{\mathcal{T}|j k}(\theta, n)$ to be
solutions of the first two equations in (\ref{kre}) for $k=2$
without poles in the physical sheet ${\rm Im}(\theta) \in
[0,\pi]$. That is,
\begin{eqnarray}
F_{\text{min}}^{\mathcal{T}|kj}(\theta,
n)=F_{\text{min}}^{\mathcal{T}|jk}(-\theta, n)S_{kj}(\theta)=
F_{\text{min}}^{\mathcal{T}|j\, k+1}(2\pi i-\theta, n) \qquad
\forall \quad j,k \label{mini}
\end{eqnarray}
where the $S$-matrix is given by (\ref{s}). Repeated use of the
above equations leads to the following constraints:
\begin{eqnarray}
  F_{\text{min}}^{\mathcal{T}|i \, i+k}(\theta,n) &=&F_{\text{min}}^{\mathcal{T}|j
  \,
  j+k}(\theta,n) \qquad
\forall \quad i,j,k\label{use}\\
F_{\text{min}}^{\mathcal{T}|1 j}(\theta,
n)&=&F_{\text{min}}^{\mathcal{T}|11}(2\pi(j-1)i-\theta, n )\qquad
\forall \quad j\neq 1.\label{use2}
\end{eqnarray}
These equations show that computing just the form factor
$F_{\text{min}}^{\mathcal{T}|11}(\theta, n)$ is enough to
determine all minimal form factors of the theory. A consequence of
these equations is that this minimal form factor must have no
poles in the extended strip $\text{Im}(\theta) \in [0, 2\pi n]$.
From the equations above it is easy to deduce
\begin{equation}
  F_{\text{min}}^{\mathcal{T}|11}(\theta, n)=F_{\text{min}}^{\mathcal{T}|11}(-\theta, n)S(\theta)=
  F_{\text{min}}^{\mathcal{T}|11}(-\theta+ 2\pi n i, n). \label{rel}
\end{equation}
In order to develop a systematic procedure to solve these
equations it is useful to recall that, for a standard local
operator the minimal form factor equations take the form
\begin{equation}
  f_{11}(\theta)=f_{11}(-\theta) {S}(n\theta)= f_{11}(-\theta+ 2\pi
  i),
   \label{rel2}
\end{equation}
provided that the $S$-matrix of the theory is given by
$S(n\theta)$.
 Thus given a solution to the previous equation, the function
$F_{\text{min}}^{\mathcal{T}|11}(\theta, n)=f_{11}(\theta/n)$ is automatically a solution of
(\ref{rel}).

\begin{figure}
\bc
\includegraphics[width=6cm,height=4cm]{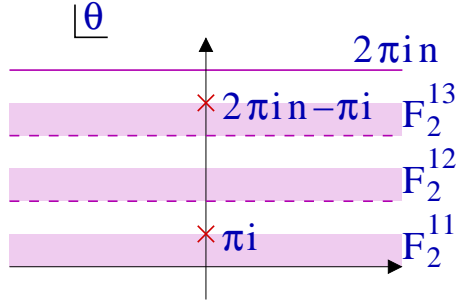}
\ec \caption{[Color online] The structure of the function
$F_2^{\tw|11}(\theta)$ in the extended sheet ${\rm
Im}(\theta)\in[0,2\pi n]$, in the case $n=3$. Crosses indicate the
positions of the kinematic singularities. Shaded regions represent
the physical sheets of the form factors $F_2^{\tw|11}(\theta)$,
$F_2^{\tw|12}(\theta)$ and $F_2^{\tw|13}(\theta)$.}
\label{fig-structure}
\end{figure}

 In the context of integrable models, a systematic way
of solving such type of equations has been developed whereby,
given an integral representation for $S(\theta)$, an integral
representation of $f_{11}(\theta)$ can be readily obtained \cite{KW}. For
diagonal theories, the integral representation of the $S$-matrix
takes the form
\begin{equation}
    S(\theta)=\exp \left[\int_{0}^{\infty} \frac{dt }{t} g(t) \sinh\left(\frac{t \theta}{i
    \pi}\right)\right], \label{irep}
\end{equation}
where $g(\theta)$ is a function which depends of the theory under
consideration. A trivial consequence of the previous equation is
\begin{equation}
    S(n \theta)=\exp\left[ \int_{0}^{\infty} \frac{dt }{t} g(t/n) \sinh\left(\frac{t \theta}{i
    \pi}\right)\right],
\end{equation}
and from here, it is easy to show that
\begin{equation}
    f_{11}(\theta)=\mathcal{N} \exp\left[\int_{0}^{\infty} \frac{dt}{t \sinh(n t)} g(t) \sin^{2}\left(\frac{i t n}{2}\left(1+\frac{i\theta}{\pi}\right)
    \right)\right]
\end{equation}
where $\mathcal{N}$ is a normalization constant. Therefore, the desired solution is
\begin{equation}
    F_{\text{min}}^{\mathcal{T}|11}(\theta,n)=
    f_{11}(\theta/n)=\mathcal{N} \exp\left[\int_{0}^{\infty} \frac{dt}{t \sinh(n t)} g(t) \sin^{2}\left(\frac{i t}{2}\left(n+\frac{i\theta}{\pi}\right)
    \right)\right].
\end{equation}
So far in this section we have computed the minimal form factors.
However what we ultimately need are the full two-particle form
factors. These are solutions of (\ref{mini}) which include poles
in the extended physical strip mentioned before. Their pole
structure is determined by the kinematical residue equations
together with (\ref{mini}). According to these equations the form
factor  $F_{2}^{\mathcal{T}|11}(\theta,n)$ has two poles in the
extended physical sheet at $\theta=  i\pi $ and $\theta=i
\pi(2n-1)$. It is not difficult to show that a  solution of
(\ref{mini}) which is consistent with the above pole structure is
given by
\begin{equation}
F_{2}^{\mathcal{T}|jk}(\theta)=\frac{ \langle \mathcal{T}\rangle
\sin\left(\frac{\pi}{n}\right)}{2 n \sinh\left(\frac{ i\pi
(2(j-k)-1)+ \theta}{2n}\right)\sinh\left(\frac{i\pi
(2(k-j)-1)-\theta}{2n}\right)}
\frac{F_{\text{min}}^{\mathcal{T}|jk}(\theta,
n)}{F_{\text{min}}^{\mathcal{T}|jk}(i \pi, n)},\label{full}
\end{equation}
where the normalization has been chosen so that the kinematical
residue equation gives
\begin{equation}
    F_{0}^{\mathcal{T}}=\langle \mathcal{T}\rangle.
\end{equation}
In addition, the constant factor $\sinh(\pi/n)$ guarantees that
all form factors vanish for $n=1$ as expected, since in that case
the field $\mathcal{T}$ can be identified with the identity.
The structure of two-particle form factors is depicted in Fig. \ref{fig-structure}.

For the field $\t\tw$, the exchange relations imply that form factors of the field
$\tilde{\mathcal{T}}$ are equal to those of the field $\mathcal{T}$ up
to the transformation $i \rightarrow n-i$ for each particle $i$. At the level of the two particle form factors, this means that
\begin{equation}
    F_2^{\mathcal{T}|ij}(\theta,n)=F_2^{\tilde{\mathcal{T}}|(n-i)
    (n-j)}(\theta,n).
\end{equation}
This property can be combined with (\ref{use})-(\ref{use2}) to
show that
\begin{eqnarray}
   F_2^{\tilde{\mathcal{T}}|11}(\theta,n)&=&F_2^{\mathcal{T}|11}(\theta,n).\\
  F_2^{\tilde{\mathcal{T}}|1j}(\theta,n)&=&F_2^{\mathcal{T}|11}(\theta+2\pi i(j-1),n).
\end{eqnarray}

\subsection{The Ising model}\indent \\

\noindent The Ising model is, together with the free Boson theory,
the simplest integrable model we can possibly consider. It
describes a real free fermion and therefore the
scattering matrix is simply
\begin{equation}
S(\theta)=-1.
\end{equation}
Form factors of local fields of the Ising model were first
computed in \cite{KW,YZam} and later on in \cite{Cardybast} for
so-called descendant fields. A solution of (\ref{mini}) for
$j=k=1$ is given by
\begin{equation}
F_{\text{min}}^{\mathcal{T}|11}(\theta)=-i\sinh\left(\frac{\theta}{2n}\right).\label{is}
\end{equation}
This is in fact the standard minimal form factor already employed
in \cite{KW,YZam}, with $\theta \rightarrow\theta/n$.

\subsection{The sinh-Gordon model}\indent \\

\noindent  The sinh-Gordon model is a quantum integrable model
possessing a single particle spectrum and no bound states. The
corresponding $S$-matrix \cite{SSG2,SSG3,SSG} is given by
   \begin{equation}\label{smatrix}
    S(\theta)=\frac{\tanh\frac{1}{2}(\theta-i \frac{\pi B}{2}) }{\tanh \frac{1}{2}(\theta+i \frac{\pi
    B}{2})}.
\end{equation}
The parameter $B \in [0,2]$ is the effective coupling constant
which is related to the coupling constant $\beta$ in the
sinh-Gordon Lagrangian \cite{toda2,toda1} as

\begin{equation}\label{BB}
    B(\beta)=\frac{2\beta^2}{8\pi + \beta^2},
\end{equation}
under CFT normalization \cite{za}. The $S$-matrix is obviously
invariant under the transformation $B\rightarrow 2-B$, a symmetry
which is also referred to as week-strong coupling duality, as it
corresponds to $B(\beta)\rightarrow B(8 \pi \beta^{-1})$ in
(\ref{BB}). The point $B=1$ is known as the self-dual point. Form
factors of the sinh-Gordon model were first computed in
\cite{FMS}. The program was thereafter extended to other operators
in \cite{KK} and more recently in \cite{Delfino:2006te}. The
$S$-matrix above admits an integral representation which is given
by (\ref{irep}), with
\begin{equation}\label{ft}
    g(t)=\frac{8\sinh\left( \frac{tB}{4}\right) \sinh\left(\frac{t}{2}\left(1-\frac{B}{2} \right)
    \right)\sinh\left( \frac{t}{2}\right)}{\sinh t}.
\end{equation}
Therefore, the minimal form factor is given by
\begin{equation}
 F_{\text{min}}^{\mathcal{T}|11}(\theta)=\exp\left[-2
 \int_{0}^{\infty} \frac{dt \sinh \frac{t B}{4} \sinh  \frac{t(2-B)}{4}}{t \sinh(n t)
 \cosh \frac{t}{2} } \cosh t\left(n+\frac{i\theta}{\pi}
    \right)\right],\label{int}
\end{equation}
where we have chosen the normalization
$\mathcal{N}=F_{\text{min}}^{\mathcal{T}|11}(i \pi n)$. Employing
the identity
\begin{equation}
    \int_{0}^{\infty} \frac{dt}{t}\frac{\sinh(\alpha t)\sinh(\beta t) e^{-\gamma
    t}}{\sinh(u t)}=\frac{1}{2}\log\left[\frac{\Gamma\left(\frac{\alpha+\beta+\gamma+u}{2u} \right)
    \Gamma\left(\frac{-\alpha-\beta+\gamma+u}{2u} \right)}{\Gamma\left(\frac{-\alpha+\beta+\gamma+u}{2u} \right)
    \Gamma\left(\frac{\alpha-\beta+\gamma+u}{2u} \right)} \right],
\end{equation}
where $\Gamma(x)$ is Euler's gamma function, we obtain the
alternative representation
\begin{eqnarray}
 \log(F_{\text{min}}^{\mathcal{T}|11}(\theta))&=&\sum_{k=0}^{\infty}(-1)^k
 \log \left[ \frac{\Gamma\left(\frac{2n-2w+B+2k}{4n} \right)\Gamma\left(\frac{2n+2w+B+2k}{4n} \right)
 }{\Gamma\left(\frac{n-w+k}{2n} \right)\Gamma\left(\frac{n+w+k}{2n}
\right)}\right.\nonumber \\
&& \qquad\qquad\quad\times \left. \frac{
 \Gamma\left(\frac{2n-2w+2-B+2k}{4n} \right)\Gamma\left(\frac{2n+2w+2-B+2k}{4n} \right)}
 {\Gamma\left(\frac{n-w+k+1}{2n} \right)\Gamma\left(\frac{n+w+k+1}{2n}
 \right)}\right]
\end{eqnarray}
with $w=n+ i \theta/\pi$, or equivalently
\begin{eqnarray}
 && F_{\text{min}}^{\mathcal{T}|11}(\theta) = \prod_{k=0}^{\infty}\left[ \frac{\Gamma\left(\frac{2n-2w+B+4k}{4n} \right)
 \Gamma\left(\frac{2n+2w+B+4k}{4n} \right)
 \Gamma\left(\frac{2n-2w+2-B+4k}{4n} \right)\Gamma\left(\frac{2n+2w+2-B+4k}{4n} \right)}
 {\Gamma\left(\frac{n-w+2k}{2n} \right)\Gamma\left(\frac{n+w+2k}{2n} \right)}\right. \nonumber \\
 &&\times  \left.\frac{\Gamma\left(\frac{n-w+2k+2}{2n} \right)
 \Gamma\left(\frac{n+w+2k+2}{2n} \right)}{\Gamma\left(\frac{2n-2w+B+4k+2}{4n} \right)
 \Gamma\left(\frac{2n+2w+B+4k+2}{4n} \right)
 \Gamma\left(\frac{2n-2w+4-B+4k}{4n} \right)\Gamma\left(\frac{2n+2w+4-B+4k}{4n} \right)}
 \right].\label{xx}
\end{eqnarray}
As a consistency check, it is quite easy to show that for $n=1$
the minimal form factor above is the standard minimal form factor
associated to local fields in the sinh-Gordon model computed in
\cite{FMS}. In appendix A we will show how the same expression can
be derived form the angular quantization scheme proposed in
\cite{freefield1} and later carried out for the exponential fields
of various models (including the sinh-Gordon model) in
\cite{freefield2}.

\section{Identifying the ultraviolet conformal dimension of  $\mathcal{{T}}$}\label{identifying}

In this section we verify that the form factors constructed above agree with the properties of the
operator $\mathcal{{T}}$ at conformal level, that is, in the ultraviolet limit. As is well-known, the
form factor program provides a way of carrying out this
verification by allowing us to compute (at least in an
approximate way) the correlation functions of various fields of an
integrable quantum field theory. In the ultraviolet limit, it is
possible to relate a particular correlation function to the holomorphic
conformal dimension of a primary field by means of the
so-called $\Delta$-sum rule:
\begin{equation}\label{delta}
    \Delta^{\tw} = \Delta^{\t\tw}=-\frac{1}{2\langle \mathcal{T}
    \rangle}\int_{0}^{\infty} r
    \left\langle \Theta(r) \t\tw(0)  \right\rangle dr
\end{equation}
(where the integration is on a space-like ray), originally
proposed by G.~Delfino, P.~Simonetti and J.L.~Cardy in \cite{DSC},
where $\Theta$ is the local operator corresponding to the trace of
the stress-energy tensor. The first equality, expected from CFT,
holds from the $\Delta$-sum rule thanks to the fact that $\Theta$
commute with $\tw$ and that $\Theta^\dag=\Theta$. The holomorphic
conformal dimension is related to the scaling dimension by $d_n =
2\Delta^\tw$, where $d_n$ is expected to be (\ref{scdim}).

By introducing a sum over all quantum states
and carrying out the $r$-integration, the expression above can be
rewritten as
\begin{eqnarray}
\Delta^{\mathcal{T}} &=&-\frac{1}{%
2\left\langle \mathcal{T}\right\rangle }\sum_{k=1}^{\infty
}\sum_{\mu _{1}\ldots \mu _{k}}\int\limits_{-\infty }^{\infty
}\ldots
\int\limits_{-\infty }^{\infty }\frac{d\theta _{1}\ldots d\theta _{k}}{%
k!(2\pi )^{k}\left( \sum_{i=1}^{k}m_{\mu _{i}}\cosh \theta
_{i}\right) ^{2}}
\nonumber \\
&&\times F_{k}^{\Theta |\mu _{1}\ldots \mu _{k}}(\theta
_{1},\ldots ,\theta _{k})\,\left( F_{k}^{\mathcal{T}|\mu
_{1}\ldots \mu _{k}}(\theta _{1},\ldots ,\theta _{k})\,\right)
^{*}\,\,\,, \label{dcorr}
\end{eqnarray}
where the sum in $\mu_i$ with $i=1,\ldots, k$ is a sum over
particle types in the theory under consideration. The sum starts
at $k=1$ since we are considering ``connected" correlation
functions, that is, the $k=0$ contribution has been subtracted.
The sum above, can only be carried out in particularly simple
cases. For most models, one must be content with evaluating just
the first few contributions to the sum. Fortunately, the many
studies carried out in the last years provide strong evidence that
the sum above is convergent and that in fact, the first few terms
provide the main contribution to the final result. Indeed, the
convergence is often so good that considering only the
contribution with $k=2$ already provides very precise results (see
e.g. \cite{FMS}). Expecting a similar behaviour also in our case,
we will approximate the sum above by the two-particle
contribution, that is
\begin{eqnarray}
\Delta^{\mathcal{T}}  \approx -\frac{n}{ 2\left\langle
\mathcal{T}\right\rangle } \int\limits_{-\infty }^{\infty
}\int\limits_{-\infty }^{\infty }\frac{d\theta _{1} d\theta _{2}
F_{2}^{\Theta |11}(\theta _{12}) F_{2}^{\mathcal{T}|11}(\theta
_{12},n)^{*}}{2 (2\pi )^{2} m^2 \left( \cosh \theta _{1} + \cosh
\theta_2 \right) ^{2}}
 . \label{dcorr2}
\end{eqnarray}
The factor of $n$ is a consequence of summing over all particle
types and using (\ref{use}). In addition, the only non-vanishing
contribution comes from form factors involving only one particle
type, since we are considering $n$ non-interacting copies of the
model. This implies that
\begin{equation}
   F_{2}^{\Theta|ij}(\theta)=0\qquad \forall \qquad i\neq j.
\end{equation}
Changing variables to $\theta=\theta_1-\theta_2$ and
$\theta'=\theta_1+\theta_2$ we obtain,
\begin{eqnarray}
\Delta^{\mathcal{T}} &\approx &-\frac{n}{%
2\left\langle \mathcal{T}\right\rangle } \int\limits_{-\infty
}^{\infty }\int\limits_{-\infty }^{\infty }\frac{d\theta d\theta'
 F_{2}^{\Theta |11}(\theta )
F_{2}^{\mathcal{T}|11}(\theta,n )^{*}}{2 (2\pi )^{2} m^2 \left(2
\cosh (\theta/2)  \cosh
(\theta'/2) \right) ^{2}}  \nonumber \\
&=&-\frac{n}{32 \pi^2 m^2 \left\langle \mathcal{T}\right\rangle }
\int\limits_{-\infty}^{\infty }d \theta\, \frac{F_{2}^{\Theta
|11}(\theta) F_{2}^{\mathcal{T}|11}(\theta, n )^{*}}{\cosh^2(
\theta/2)}.
\end{eqnarray}
Let us now evaluate this integral both for the Ising and
sinh-Gordon models.

\subsection{The Ising model} \indent \\

\noindent For the Ising model the only non-vanishing form factor
of the trace of the stress-energy tensor is the $2$-particle form
factor. Hence, the two-particle approximation (\ref{dcorr2})
becomes exact. The two-particle form factors are given by
\begin{equation}
F_{2}^{\mathcal{T}|11}(\theta)=\frac{-i  \langle
\mathcal{T}\rangle \cos\left(\frac{\pi}{2n}\right)}{ n
\sinh\left(\frac{ i\pi +
\theta}{2n}\right)\sinh\left(\frac{i\pi-\theta}{2n}\right)}
{\sinh\left(\frac{\theta}{2n}\right)},\qquad
F_{2}^{{\Theta}|11}(\theta)= -2 \pi i m^2
\sinh\left(\frac{\theta}{2}\right),
\end{equation}
and therefore
\begin{eqnarray}
\Delta^{\mathcal{T}} = -\frac{1}{16 \pi} \int\limits_{-\infty
}^{\infty }\frac{
\cos\left(\frac{\pi}{2n}\right)\sinh\left(\frac{\theta}{2n}\right)
\sinh\left(\frac{\theta}{2}\right)}{ \sinh\left(\frac{ i\pi +
\theta}{2n}\right)\sinh\left(\frac{i\pi-\theta}{2n}\right)\cosh^2\left(\frac{\theta}{2}\right)}\,
d \theta .
\end{eqnarray}
It is easy to check numerically that the above integral exactly
reproduces the expected value (\ref{scdim}) for $c=1/2$,
\begin{equation}\label{quiteknut}
    2\Delta^{\mathcal{T}}=\frac{1}{24}\left(n-\frac{1}{n}\right) = d_n,
\end{equation}
for any value of $n$. The integral can also be computed
analytically, at least for $n$ even. In this case, shifting $t$ by
$2 \pi n i$ the integral above changes by a sign and therefore it
is possible to show
\begin{equation}
    2\Delta^{\mathcal{T}}=2 \pi i \sum_{j=1}^{n} r_j,
\end{equation}
where $r_j$ are the residues of the poles of the integrand at $t=i
\pi (2j-1)$, with $j=1,\ldots,n$. Of those, the poles at $j=1, n$
are triple whereas all the others are double poles. A tedious but
straightforward computation yields
\begin{eqnarray}
   2\pi i r_1&=&\frac{ -1 + n^2 + 6\cot (\frac{\pi }{n})/\sin (\frac{\pi }{n}) }{48 n
   },\nonumber \\
    2 \pi i r_j &=& \frac{ {\left( -1 \right) }^{j+1}\left( \cot (\frac{\left( j-1 \right)\pi }{n})/\sin (\frac{\left( j-1 \right) \pi }{n}) +
        \cot (\frac{j\pi }{n})/\sin (\frac{j\pi }{n}) \right)}{8 n}, \quad 1<j<n,\nonumber \\
 2\pi i r_n &=&
 \frac{(-1)^{n+1}\left( 4 - n^2 - 3\cot^2 (\frac{\pi }{2 n}) + 6 \cot^2 (\frac{\pi }{n}) \right) }{48
 n}.
\end{eqnarray}
Finally, we need to add up all these residues. The sum over the
$r_j$ residues becomes in fact very simple, since it is a
telescopic series. We obtain,
\begin{equation}
  2 \pi i  \sum_{j=2}^{n-1}
  r_j=-\frac{\cot(\frac{\pi}{n})}{4n\sin(\frac{\pi}{n})},
\end{equation}
which gives (\ref{quiteknut}).

\subsection{The sinh-Gordon model}\indent \\

\noindent In this case, the relevant $2$-particle form factors are
given by
\begin{equation}
F_{2}^{\mathcal{T}|11}(\theta)=\frac{ \langle \mathcal{T}\rangle
\sinh\left(\frac{\pi}{n}\right)}{2 n \sinh\left(\frac{ i\pi +
\theta}{2n}\right)\sinh\left(\frac{i\pi-\theta}{2n}\right)}
\frac{F_{\text{min}}^{\mathcal{T}|11}(\theta,
n)}{F_{\text{min}}^{\mathcal{T}|11}(i \pi, n)},\qquad
F_{2}^{{\Theta}|11}(\theta)= 2 \pi m^2
\frac{F_{\text{min}}^{\mathcal{T}|11}(\theta,
1)}{F_{\text{min}}^{\mathcal{T}|11}(i \pi, 1)}.
\end{equation}
The form factors of $\Theta$ were computed in \cite{FMS}. Since
$\Theta$ is a local operator, its minimal form factor is given by
(\ref{xx}) with $n=1$. Thus,
\begin{eqnarray}
\Delta^{\mathcal{T}} \approx -\frac{1}{32 \pi}
\int\limits_{-\infty }^{\infty }\frac{
\sin\left(\frac{\pi}{n}\right)}{ \sinh\left(\frac{ i\pi +
\theta}{2n}\right)\sinh\left(\frac{i\pi-\theta}{2n}\right)\cosh^2\left(\frac{\theta}{2}\right)}
\frac{F_{\text{min}}^{\mathcal{T}|11}(\theta,
n)^*}{F_{\text{min}}^{\mathcal{T}|11}(i \pi, n)^*}
\frac{F_{\text{min}}^{\mathcal{T}|11}(\theta,
1)}{F_{\text{min}}^{\mathcal{T}|11}(i \pi, 1)} d\theta.
\end{eqnarray}
The tables below show the result of carrying out this integral
numerically for various values of $n$ and $B$. Next to each value
of $n$ in brackets we show for reference the expected value of
$\Delta^{\mathcal{T}}$, as predicted by the CFT formula (\ref{scdim}) (with, again, $\Delta^\tw = d_n/2$).
\begin{center}
\begin{tabular}{|l|l|l|l|l|}
\hline & $n=2$ (0.0625) & $n=3$ (0.1111) & $n=4$ (0.1563) & $n=5$ (0.2)\\
\hline
 $B=0.02$& 0.0620& 0.1114& 0.1567  & 0.2007 \\ \hline
 $B=0.2$ & 0.0636 & 0.1135& 0.1599  & 0.2048  \\\hline
 $B=0.4$ & 0.0636& 0.1148 & 0.1620   & 0.2074   \\\hline
 $B=0.6$ & 0.0643&0.1155 & 0.1631   &0.2088   \\\hline
 $B=0.8$ & 0.0644&0.1158 & 0.1636  & 0.2096 \\\hline
 $B=1$  & 0.0644& 0.1159 & 0.1637 &0.2098   \\\hline
\end{tabular}
\end{center}
\begin{center}
\begin{tabular}{|l|l|l|l|l|l|}
\hline & $n=6$ (0.2431)& $n=7$ (0.2857)& $n=8$ (0.3281)& $n=9$ (0.3704) & $n=10$ (0.4125)\\
\hline
 $B=0.02$ & 0.2436   &  0.2864 & 0.3289 & 0.3712  &0.4135  \\ \hline
 $B=0.2$ & 0.2488  & 0.2925  &0.3360   & 0.3793  & 0.4225 \\\hline
 $B=0.4$ & 0.2522  &0.2966   & 0.3407  & 0.3846 & 0.4284 \\\hline
 $B=0.6$ & 0.2540 & 0.2988  &  0.3433 & 0.3876  & 0.4317 \\\hline
 $B=0.8$ &0.2550  & 0.2999  & 0.3446 & 0.3890 & 0.4334 \\\hline
 $B=1 $ & 0.2552  & 0.3002   & 0.3449  & 0.3895 & 0.4339\\\hline
\end{tabular}
\end{center}
The figures obtained are extremely close to their expected value
for all choices of $B$ and $n$. In most cases they are slightly
above the expected value. This is not surprising since the 4- and
higher particle contributions are not necessarily positive.

\section{Two-point functions and the entanglement entropy}
\label{sectentang}

As explained before, the entanglement entropy is given by the
derivative with respect to $n$ of the two-point function $\langle
\mathcal{T}(r) \tilde{\mathcal{T}}(0)\rangle$ evaluated at $n=1$.
The behaviour of the entropy at short separations $r\ll m^{-1}$ is
described by the conformal limit of the model and is already well
known \cite{Calabrese:2004eu,Calabrese:2005in}. At large
separations $r\gg m^{-1}$ (in the infrared limit) it is also known
to saturate. Here we would like to evaluate the first correction
to the entropy in the infrared limit. In this limit, the
two-particle contribution provides the first sub-leading
exponential term in the correlation function $\langle
\mathcal{T}(r) \tilde{\mathcal{T}}(0)\rangle$.
 Hence, the two-particle approximation should provide both the saturation value of the entropy, coming from the
 disconnected part of the correlation function, and the exact first exponential correction at large $rm$,
 coming from the two-particle contributions.

The two-point function in the two-particle approximation is given by
\begin{eqnarray}
\langle \mathcal{T}(r) \tilde{\mathcal{T}}(0)\rangle &\approx&
\langle\mathcal{T}\rangle^2+ \sum_{i,j=1}^n\int\limits_{-\infty
}^{\infty } \int\limits_{-\infty }^{\infty }\frac{d\theta
_{1}d\theta _{2}}{ 2!(2\pi )^{2}}\left|F_{2}^{\mathcal{T} |ij}(\theta
_{12},n)\right|^2\,e^{-rm
(\cosh\theta_1 + \cosh \theta_2)}\nonumber\\
&=&\langle\mathcal{T}\rangle^2\left(1 + \frac{n}{4
\pi^2}\int\limits_{-\infty }^{\infty } d\theta f(\theta,n)
K_{0}(2rm\cosh(\theta/2))\right),\label{ent1}
\end{eqnarray}
where we changed variables as in section \ref{identifying}, $K_0(z)$
is the Bessel function resulting from carrying out one of the
integrals, and we defined
\begin{eqnarray}
 {\langle\mathcal{T}\rangle^{2}}f(\theta,n) &=& \sum_{j=1}^{n}\lt|F_{2}^{\mathcal{T} |1j}(\theta,n)\rt|^2 \label{ftn} \\
   &=& \left|F_{2}^{\mathcal{T} |11}(\theta,n) \right|^2 +
   \sum_{j=1}^{n-1} \lt|F_{2}^{\mathcal{T} |11}(-\theta+2\pi ij,n)\rt|^2\nonumber
\end{eqnarray}
(with $f(\theta,1)=0$).
Notice that the function above is only defined for integer values of $n$.

In order to obtain the entropy we should now analytically continue
the two-point function (\ref{ent1}), as function of $rm$ and $n$,
from $n=1,2,3,\ldots$ to $n\in[1,\infty)$, compute the derivative
with respect to $n$ and evaluate the result at $n=1$. The analytic
continuation is of course not unique. We will choose the one which
is such that for ${\rm Re}(n)>0$, the two-point function divided by $\bra\tw\ket^2$
is $O(e^{q n})$ as $n\to\infty$ for some $q<\pi$. It is then unique by Carlson's
theorem \cite{Rubel55}. This choice is motivated by the fact that the trace
(\ref{thetrace}) has this behavior with $q<0$ for any finite
system, since the eigenvalues of $\rho_A$ are real an positive,
and are normalised to sum to 1. Of course, this is merely a
motivation. For infinite systems, eigenvalues should have dense
components and we could have algebraic behaviors; the scaling
limit and the limit $n\to\infty$ may not commute, so that the
coefficient of the small-distance power-law $\ep^{2d_n}$ may be
divergent; and the limit $rm\to\infty$ and $n\to\infty$ of this
coefficient may also not commute, so that the behavior in $n$ of
the large-$rm$ expansion coefficients may also be divergent. We
expect that this gives at most algebraic divergencies in $n$, but
a better understanding would be desirable.

There are three main observations necessary to understand the
analytic continuation of (\ref{ent1}) and the evaluation of the
derivative at $n=1$:
\begin{itemize}
\item {\em Structure of the analytic continuation:} There is no natural
(as described above) analytic continuation of $f(0,n)$ from
$n=1,2,3,\ldots$ to $n\in[1,\infty]$. Instead, there is such an analytic
continuation from $n=2,3,4,\ldots$ to $n\in[1,\infty]$, which we will
denote by $\t{f}(n)$, but it has the property that
\begin{equation}
   \tilde{f}(1) \neq f(0,1) = 0\,.
\end{equation}
On the other hand, the function $f(\theta,n)$ does not have any such feature for $\theta\neq0$: its analytic continuation $\t{f}(\theta,n)$ to $n\in[1,\infty]$ agrees with $f(\theta,n)$ for all $n=1,2,3,\ldots$. In particular, $\t{f}(\theta,1)=0$. Hence, as $n\to1$, the function $\t{f}(\theta,n)$ does not converge uniformly on $\theta\in(-\infty,\infty)$.
\item {\em Kinematic singularities:} The non-uniform convergence of $\t{f}(\theta,n)$ can be seen to be a consequence of the collision of the kinematic singularities of form factors $F_2^{\tw|11}(\theta,n)$ (at $\theta=i\pi$ and $\theta = i\pi (2n-1)$) that occur when $n\to1$ (see Fig. \ref{fig-collide}). The idea is made clear from considering Poisson's resummation formula. Consider a sum of the type $\sum_{j=1}^{n-1} s(\theta,j)$, as in the r.h.s. of (\ref{ftn}), re-written using Poisson's resummation formula, which holds whenever $s(\theta,n)=s(\theta,0)$ (the summand $s(\theta,j)$ in (\ref{ftn}) indeed satisfies this by unitarity):
\beq
    \sum_{j=1}^{n-1} s(\theta,j) = \sum_{k=-\infty}^\infty (s_{nk}-s_k)~,\quad     s_k = \int_0^{n} dj\, e^{ -\frc{2\pi i jk}n} s(\theta,j)~.
\eeq
For any $\theta\neq0$, no singularity of $s(\theta,j)$ occur on the integration path\footnote{The sum over $k$ can be made absolutely convergent by a slight imaginary shift of the $j$-integration path defining $s_k$. Since $s(\theta,j)$ is in fact not periodic in $j$, there are contributions along ${\rm Re}(j)=0,n$ additionally to the shifted path, with coefficients that vanish linearly as ${\rm Im}(j)\to0$. This linear vanishing guarantees a vanishing like $1/k^2$ of $s_k$ at large $k$, making the sum convergent.} defining $s_k$, so that when $n\to1$, the result should vanish. In fact, since $s(\theta,j)$ itself vanishes like $(n-1)^2$ for $\theta\neq0$, the result vanishes like $(n-1)^3$. When $\theta\to0$, the poles of $s(\theta,j)$, which are at $j=\frc12\pm\frc{\theta}{2\pi i}$ and $j=n-\frc12\pm\frc{\theta}{2\pi i}$, pinch the integration path, but the quantities $s_k$ are still finite because the divergent contributions at ${\rm Re}(j)=\frc12$ and ${\rm Re}(j)=n-\frc12$ cancel out -- we are left with principal-value integrals with two double-poles (and the resulting conditionally convergent sum over $k$ has a unique finite value defined by the limit $\theta\to0$). However, since these two double poles collide at $n=1$ and fuse into a higher order pole, there is no guarantee that the result vanishes as $n\to1$.
\begin{figure}
\bc
\includegraphics[width=14cm,height=4cm]{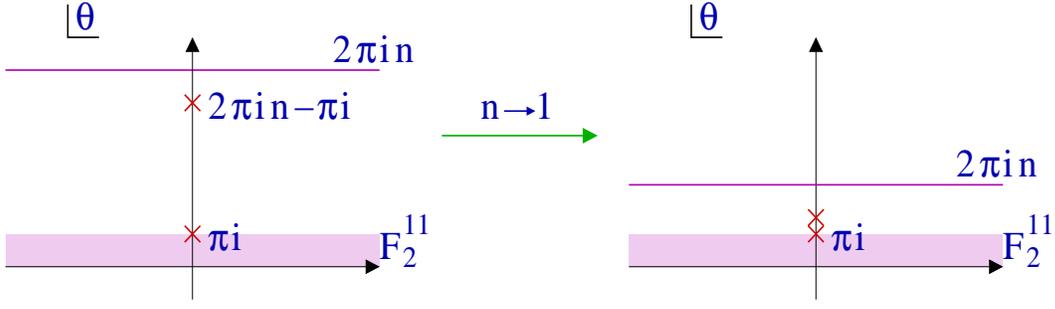}
\ec \caption{[Color online] The collision of kinematic
singularities when $n\to1$.} \label{fig-collide}
\end{figure}
\item {\em delta-function:} From the arguments above, the full contributions of the derivative with respect to $n$ of the two-particle contribution to the two-point function, in the limit $n\to1$, is obtained from the region $\theta\sim0$, and is due to the kinematic singularities. In fact, notice that as $n\to1$, the derivative with respect to $n$ of $\t{f}(\theta,n)$ ``around'' $\theta=0$ should diverge: indeed, the values of $\t{f}(\theta,n)$ at $\theta=0$ and at its neighboring points should be about the same for all $n>1$, by continuity, but they should reach a finite separation at $n=1$; hence variations must be very strong near to $n=1$. This leads to the expectation that $\frc{\p}{\p n}\t{f}(\theta,n)$ at $n=1$ is proportional to $\delta(\theta)$. The contribution of the kinematic singularities to the sum in the function $\t{f}(\theta,n)$ (\ref{ftn}) is obtained from the singular behavior in $j$ of the summand $s(\theta,j)= F_{2}^{\mathcal{T} |11}(-\theta+2\pi i j,n) \lt(F_{2}^{\mathcal{T} |11}\rt)^*(-\theta-2\pi i j,n)$: \[
    s(\theta,j)\sim \frc{i\;F_{2}^{\mathcal{T} |11}(-2\theta+2\pi i n-i\pi)}{-\theta-2\pi i j+2\pi i n-i\pi}
        - \frc{i\;F_{2}^{\mathcal{T} |11}(-2\theta+i\pi)}{-\theta-2\pi i j+i\pi} + {\rm c.c.}
\]
(where ${\rm c.c.}$ means complex conjugate, for real $\theta$). It is a simple matter to perform on this expression the sum $\sum_{j=1}^{n-1}$, giving
\beqa
    \sum_{j=1}^{n-1} s(\theta,j) &\sim& \frc1{2\pi} \lt(\psi\lt(-\frc12+n+\frc{i\theta}{2\pi}\rt) - \psi\lt(\frc12+\frc{i\theta}{2\pi}\rt)\rt)
        F_{2}^{\mathcal{T} |11}(-2\theta+2\pi i n-i\pi) \n &&
        + \frc1{2\pi} \lt(\psi\lt(-\frc12+n-\frc{i\theta}{2\pi}\rt) - \psi\lt(\frc12-\frc{i\theta}{2\pi}\rt)\rt)
        F_{2}^{\mathcal{T} |11}(-2\theta+i\pi) \n && + {\rm c.c.} \no
\eeqa where $\psi(z) = d\log \Gamma(z) /dz$ is the derivative  of
the logarithm of Euler's Gamma function. This has no poles at
$\theta=0$, as the kinematic poles of the form factors involved
cancel out. The poles that are nearest to ${\rm Re}(\theta)=0$ as
$n\to1$ are at $\theta = \pm i\pi(n-1)$, coming from the form
factors involved. The residues to first order in $n-1$ give:
\beq\label{tfnm}
    \t{f}(\theta,n) \sim \t{f}(1) \lt( \frc{i\pi(n-1) }{2(\theta+i\pi(n-1))} - \frc{i\pi(n-1) }{2(\theta-i\pi(n-1))}\rt) \quad (n\to1)
\eeq
with
\beq
    \t{f}(1) = \frc12~.
\eeq This has simple poles at $\theta=\pm i\pi(n-1)$ with residues
that vanish at $n=1$, gives $\t{f}(1)$ at $\theta=0$  and vanishes
like $(n-1)^2$ as $n\to1$ for $\theta\neq0$. The limit $n\to1$, as
a distribution on $\theta$, is easily evaluated:
\beq
    \lt(\frc{\p}{\p n} \t{f}(\theta,n)\rt)_{n=1} = \pi^2 \t{f}(1) \delta(\theta)~.
\eeq
 In Appendix \ref{appf}, we give the full form of
$\t{f}(\theta,n)$ and verify that this is correct. Note that for
the free case our result is in agreement with the $n \rightarrow
1$ limit evaluated in \cite{casini1}.
\end{itemize}

Inserting this inside (\ref{ent1}) gives, using (\ref{rhon}) and
(\ref{SA}),  the entanglement entropy $S_A(rm)$ for $A$ an
interval of length $r$:
\begin{equation}
    S_A(rm)=-\frac{c}{3}\log(\epsilon m) + U - \frc{1}{8} K_0(2rm) + O\lt(e^{-4rm}\rt)
\end{equation}
where
\beq
    U = - \frc{d}{dn} \lt(  m^{-2d_n}\bra \tw\ket^2\rt)_{n=1}~.
\eeq
Therefore the sub-leading large $rm$ terms in the entropy are given by the Bessel function $K_0(2rm)$, up to terms that are exponentially smaller (coming from the neglected 4-particle form factors). The term $U$ involving the derivative of the vacuum expectation value of $\tw$ has a universal meaning since the normalisation of $\tw$ has been fixed (see (\ref{normalisationtw})). More precisely, with this normalisation, the entanglement entropy at short interval length is $S_A(rm) = -\frac{c}{3}\log(\ep/r)+O(rm)$. It is of course possible to define the quantity $U$ in a way that is obviously universal, valid for any choice of short-distance cutoff:
\beq
    U = \lim_{\xi\to\infty} \lt(S_A(\xi) - S_A(\xi^{-1}) - \frc{c}3 \log\xi\rt).
\eeq

We will now proceed to evaluate explicitly the functions $f(0,n)$ and $\tilde{f}(n)$ for the Ising and sinh-Gordon models, verifying some of the results above.
It is worth noting that the function $f(0,n)$, which we study in more detail below, also has a meaning as the coefficient of the leading exponential correction to the partition function on the Riemann surface $\orb_{n,0,r}$:
\beq
    \langle \mathcal{T}(r) \tilde{\mathcal{T}}(0)\rangle = \bra \tw\ket^2\lt(1+\frc{n f(0,n) e^{-2rm}}{4\pi rm} + O\lt(\frc{e^{-2rm}}{(rm)^2}\rt)\rt)
\eeq

\subsection{The Ising model}\indent \\

\noindent For the Ising model, the function (\ref{ftn}) at $\theta=0$ is given by
\begin{equation}
    f(0,n)=\frac{\cos^2\left(\frac{\pi}{2n}\right)}{n^2}\sum_{j=2}^{n}\frac{\sin^2\left(\frac{ (j-1)\pi}{n}\right)  }{
\sin^2\left(\frac{(2j-1)\pi
}{2n}\right)\sin^2\left(\frac{(2j-3)\pi}{2n}\right)}=\frac{1}{2}.\label{sum}
\end{equation}
The result of the sum can be obtained analytically as in appendix \ref{appf}. In
this case,  the integral part of the formula (\ref{tfn}) is zero,
so that $\t{f}(n)=1/2$ for all $n$. That is, the connected part of
the correlation function at large $rm$ behaves linearly with $n$,
for all values of $n$. Hence, the entanglement entropy can be
computed to
\begin{equation}
   S_A(rm) = -\frac{1}{6}\log(\epsilon m) + U_{{\rm Ising}}-\frc18 K_0(2rm) + O\lt(e^{-4rm}\rt)~.
\end{equation}
The constant $U_{{\rm Ising}}$ can also be evaluated explicitly
using the relation between the Ising model and  the free Dirac
fermionic model, as explained in Appendix \ref{appising}. We find
that the expectation value of the branch-point twist field is
\beq\label{vevtw}
    \bra \tw\ket = \lt(\frc{m}2\rt)^{\frc1{24}\lt(n-\frc1n\rt)}
     \exp\lt[\int_0^\infty \frc{dt}{4t}\lt( \frc1{\sinh t \sinh\frc{t}n} -\frc{n}{\sinh^2 t} - \frc{e^{-2t}}6 \lt(n-\frc1n\rt)\rt)\rt]
\eeq
which gives
\beq
    U_{{\rm Ising}} = \frc16 \log 2 - \int_0^\infty
    \frc{dt}{2t}\lt( \frc{t\cosh t}{\sinh^3 t} - \frc1{\sinh^2t} - \frc{e^{-2t}}3\rt) = -0.131984...
    \label{uising}
\eeq Once the value of $ U_{{\rm Ising}}$ has been fixed we can
compare our expressions for the entropy in the deep-infrared and
deep-ultraviolet regimes to existing results in the literature for the quantum Ising chain.
The comparison goes as follows: from
\cite{Calabrese:2005in,Peschel} it is possible to obtain the
expression of the entropy at large separations $r \gg m^{-1}$ in
terms of the lattice spacing $a$
\begin{equation}\label{speschel}
    S_A=-\frac{1}{6}\log(am)+ \frac{1}{2}\log 2.
\end{equation}
By comparing (\ref{speschel}) to our formula
\begin{equation}
    S_A=-\frac{1}{6}\log(\epsilon m)+   U_{{\rm Ising}},
\end{equation}
we obtain the precise relationship between the short-distance
cutoff $\epsilon$ and the lattice spacing $a$. That is
\begin{equation}\label{epsa}
    \epsilon= \frac{a}{8} e^{6 U_{{\rm Ising}}}=(0.0566227...) a.
\end{equation}

As mentioned above, formula (\ref{speschel}) follows from the
results in \cite{Calabrese:2005in,Peschel}. However this is not
entirely trivial and some clarifications are due here. First, the
formulae given in these publications are expressed in terms of a
parameter $k$ rather than $a$. The parameter $k$ is related to the
value of the transverse magnetic field $h$ of the quantum Ising chain as follows:
\begin{equation}\label{kh}
    k=\left\{%
\begin{array}{ll}
    h & \text{for } \quad h<1 \\
    h^{-1}, & \text{for } \quad h>1 \\
\end{array}%
\right.
\end{equation}
where $h=1$ corresponds to the critical point. The entropy in both
regions of values of $k$ was computed in \cite{Peschel} whereas in
\cite{Calabrese:2005in} only the $h<1$ regime was considered.
However, it is easy to show that both regimes give the same
infrared value of the entropy when $h$ approaches 1. Second, it is a
standard result that $\pm am =1-h$, where the positive sign
corresponds to $h<1$ and the negative sign corresponds to $h>1$.
Therefore, the fact that the infrared value of the entropy is the
same both for $h>1$ and $h<1$ is in agreement with what we expect
from QFT, as going from one regime to the other amounts formally
to a change in the sign of the mass $m$ which has no effect on the
value of the energy density. That is, the two regions should be
described by the same QFT in the scaling limit. The relations
between $k$ and $h$ and between $a m$ and $h$ given above allow us
to relate the parameter $k$ (in terms of which the entropy is
expressed) to the lattice spacing $a$. Once this is done we only
need to expand the expressions in \cite{Calabrese:2005in,Peschel}
around the value $k=1$. Since these formulae are given in terms of
the complete elliptic integral of the first kind $K(k)$, we need to employ the standard expansion
\begin{equation}\label{elliptic}
    K(k) \sim  -1/2 \log(1-k) + 3/2 \log
    2\quad\text{for}\quad k \to 1,
\end{equation}
to obtain (\ref{speschel}).

\begin{figure}
\bc \includegraphics[width=11cm,height=7cm,angle=0]{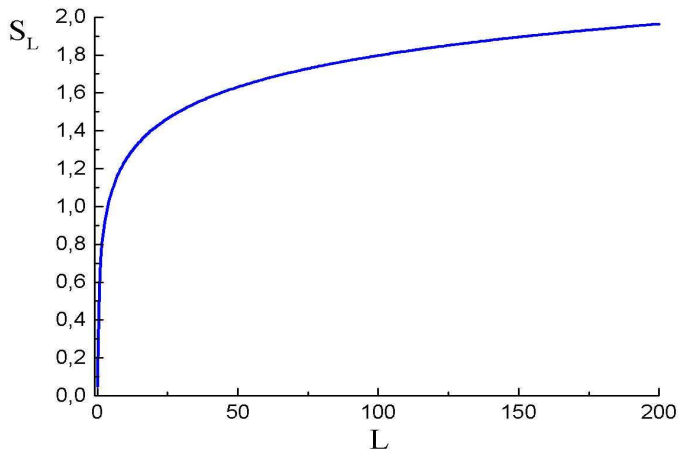}\ec
    \caption{[Color online] A plot of the function $S_{L}=\frac{1}{6}\log_{2}(L)+ 0.690413$ for $L \leq 200$. The graph is in very good agreement
    with the numerical values plotted in Fig.~9 of \cite{Latorre2}.}\label{sl}
\end{figure}

Plugging (\ref{epsa}) into  the short-distance expression of the
entropy we obtain:
\begin{equation} \label{salog}
    S_{A}=-\frac{1}{6}\log({\epsilon}/{r}) + O(rm) =-\frac{1}{6}\log({a}/{r})+0.478558... + O(rm) ~.
\end{equation}
We can compare this result to the numerical values obtained in
\cite{Latorre2} for the Ising spin chain. However, we must notice
first that in there the entropy was defined as \beq
    S_L = -\Tr_{{\cal H}_A} (\rho_A \log_2(\rho_A))~,
\eeq whereas in this paper we have used the definition
(\ref{entropy}). This means that in order to compare (\ref{salog})
to the results obtained in \cite{Latorre2} we must divide our
formula by a $\log 2$ factor. This gives
\begin{equation} \label{salog2}
    S_{L}=\frac{1}{6}\log_{2}(L)+ 0.690413...
\end{equation}
where we now introduced the parameter $L=r/a$ which is the number
of sites in the interval $A$. A plot of this function is presented
in Fig.~\ref{sl} which  is to be compared to the blue curve in
Fig.~9 of \cite{Latorre2}. Very good agreement is found, which
supports the twist-field realization proposed in appendix
\ref{appising}.

It is worth noticing that in the free case there are alternative
ways of computing the entanglement entropy
\cite{chung,Peschel2,casini3}.

\subsection{The sinh-Gordon model}\indent \\

\begin{figure}
\bc \includegraphics[width=11cm,height=7cm,angle=0]{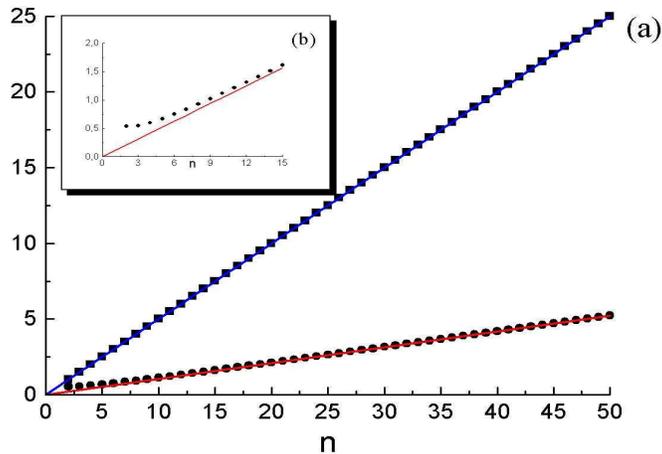}\ec
    \caption{[Color online] (a) shows 4 functions: the points are the function $n f(0,n)$ for integer
values of $n$ in the interval $[2,50]$ both for the Ising (black
squares) and sinh-Gordon (black circles) models, evaluated
numerically. The solid blue line gives the corresponding analytic
continuation $n \tilde{f}(n)$ for real values of $n$ in the
interval $[0,50]$ for the Ising model, that is the function $n/2$.
Finally the solid red line gives the function $n
\tilde{f}(\infty)$ for the sinh-Gordon model, that is a straight
line passing through the origin which describes the asymptotic
behaviour of the function $n \tilde{f}(n)$ for $n$ large. In the
sinh-Gordon case, all functions have been computed for $B=0.5$.
(b) is a magnification of the lower left corner of the sinh-Gordon
part of (a).} \label{fig-fn}
\end{figure}

\noindent In this section we will verify that $\tilde{f}(1)=1/2$
also for the sinh-Gordon model. Here, the integral part of
(\ref{tfn}) is not vanishing, hence in contrast to the Ising model
result  (\ref{sum}), $\t{f}(n)$ is in general not constant with $n$.
This can be easily seen from Fig. \ref{fig-fn} where the functions
$n f(0,n)$ and $n \tilde{f}(\infty)$ are presented, both for the
Ising and sinh-Gordon models.

The non-linearity of the function $n
\tilde{f}(n)$ in the sinh-Gordon model can be seen more clearly by magnifying the
lower left corner of Fig. \ref{fig-fn} (a), which we have done in Fig. \ref{fig-fn} (b).
Figure (b) also shows how $\tilde{f}(1)$ appears to tend to the
value $1/2$ and how the function $n \tilde{f}(n)$ deviates from
the straight line for small values of $n$. Both for the Ising and
sinh-Gordon model the figures show clearly that the function $n
f(0,n)$ has a jump at $n=1$, since $f(0,1)=0$. We will now attempt
to provide a more rigorous description of these behaviours.

In principle we would only need to perform the integral in (\ref{tfn}) to obtain $\tilde{f}(n)$ for
any values of $n$. However this is highly non-trivial due to the complexity of the minimal form factor,
as a function of $n$. We choose therefore to proceed in a different and more instructive way: we will
instead find the natural analytic continuation of the function $n f(0,n)$ numerically, as a large $n$ expansion in
powers of $1/n$.

First of all, it is an interesting exercise to try to determine
analytically the precise slope of the line $n \tilde{f}(\infty)$
in the sinh-Gordon model. This is a relatively tedious but
straightforward computation which we present in appendix
\ref{finf}. The main result is the value of $\tilde{f}(\infty)$ as
a function of $B$, which is given by
\begin{eqnarray}
   \tilde{f}(\infty)&=& \frac{8192}{\pi^2\left(4-{B}\right)^2\left(2+{B}\right)^2}
   \left[\frac{\Gamma\left(\alpha\right)\Gamma\left(\beta\right)}
   {\Gamma\left(\kappa-1\right)
  \Gamma\left(\sigma-1\right)
  }\right]^4 \left[ \frac{1}{9}  \,_{7}F_{6}\left[ \begin{array}{c}
   \frac{1}{2}, \frac{1}{2}, 1, \alpha,\alpha,\beta,\beta; 1 \\
     \frac{5}{2},\frac{5}{2},\kappa,\kappa,\sigma,\sigma \\ \end{array}\right]\right.\nonumber\\
 & + & \left. \frac{1}{75} \left(\frac{\alpha\beta}{\kappa\sigma}\right)^2 \left( \,_{7}F_{6}\left[ \begin{array}{c}
    \frac{3}{2}, \frac{3}{2}, 2, \alpha+1,\alpha+1,\beta+1,\beta+1; 1 \\
     \frac{7}{2},\frac{7}{2},\kappa+1,\kappa+1,\sigma+1,\sigma+1 \\
     \end{array}\right] \right.\right. \nonumber \\
     &+ & \left .  \left .\frac{6(8+B)^2(B-10)^2}{49(10+B)^2(B-12)^2}\,_{7}F_{6}\left[ \begin{array}{c}
    \frac{5}{2}, \frac{5}{2}, 3, \alpha+2,\alpha+2,\beta+2,\beta+2; 1 \\
     \frac{9}{2},\frac{9}{2},\kappa+2,\kappa+2,\sigma+2,\sigma+2 \\
     \end{array}\right]\right)\right],\label{slope2}
\end{eqnarray}
in terms of the generalized hypergeometric functions defined in
(\ref{fpq}) and the variables
\begin{equation}
    \alpha=\frac{3}{2}-\frac{B}{4},\quad \beta=1+\frac{B}{4},\quad
    \kappa=2-\frac{B}{4},\quad \sigma=\frac{3}{2}+\frac{B}{4}.
\end{equation}
\begin{figure}
\begin{center}
   \includegraphics[width=12cm,height=8cm,angle=0]{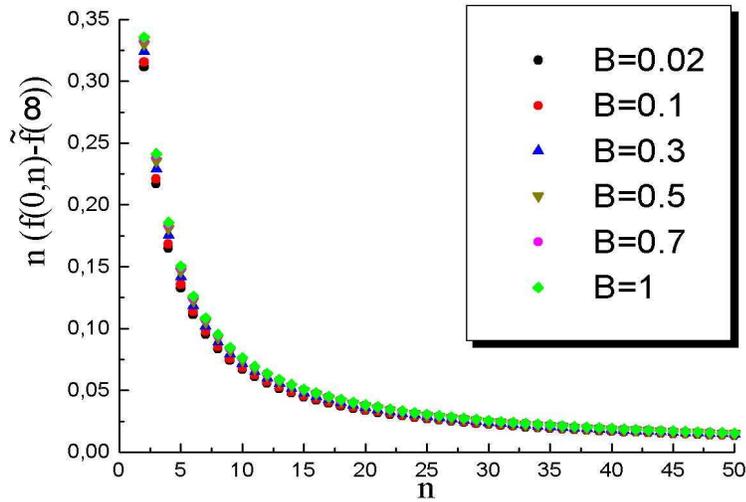}
   \end{center}
    \caption{[Color online] The function $n (f(0,n) - \t{f}(\infty))$ for integer $n$ in the
interval $[2,50]$, for various values of $B$.}
\label{fig-entropy1}
\end{figure}
As mentioned before, the data in Fig. \ref{fig-fn} (b) also
exhibit a clear deviation from the linear behaviour for small
values of $n$, which is the region we are most interested in. Fig.
\ref{fig-entropy1} shows the values of the function $n f(0,n)$ for
integer $n$ in the interval $[2,50]$ with the linear part, that is
$n \tilde{f}(\infty)$ subtracted, for various values of $B$. By
subtracting the linear part, the behaviour of the function for
large $n$ becomes more clear and appears to be dominated by a term
proportional to $1/n$. This dependence can be made more precise by
numerically fitting the various functions plotted above to a
function of the generic form
\begin{equation}
   p_{\text{fit}}(n)=
   \frac{a_0}{n}+\frac{a_1}{n^3}+\frac{a_2}{n^5}+\frac{a_3}{n^7}\,. \label{fit}
\end{equation}
The fact that only odd powers of $n$ appear is a numerical
observation. The table below contains the values of the constants
$a_0, a_1, a_2$ and $a_3$ obtained numerically for various values
of $B$, as well as the exact value of $\tilde{f}(\infty)$ from
equation (\ref{slope2}):
\begin{center}
\begin{tabular}{|l|c|c|c|c|c|}
  \hline
    $ B $ &$\tilde{f}(\infty)$& $a_0$ & $a_1$ & $a_2$ & $a_3$ \\
  \hline
  $0.02$ &$0.0952$ & $0.67(1)$ & $-0.17(2)$ &$ -0.06(1)$ &$ -0.02(3)$\\
  $0.1$ & $0.0972$& $0.68(5)$ & $-0.20(5)$ & $-0.04(8)$ & $-0.02(3)$  \\
  $0.2$ & $0.0994$& $0.70(3)$ & $-0.24(2)$ & $-0.02(9)$ & $-0.02(7)$  \\
  $0.3$ & $0.1013$& $0.71(8)$ & $-0.27(7)$ & $-0.00(9)$ & $-0.03(1)$  \\
  $0.4$ & $0.1030 $& $0.73(1)$ & $-0.30(8)$ & $0.01(0)$ & $-0.03(8)$  \\
  $0.5$ & $0.1044$& $0.74(2)$ & $-0.33(5)$ & $ 0.02(9)$ & $-0.04(4)$  \\
  $0.6$ & $0.1055$& $0.75(1)$ & $-0.35(8)$ & $0.04(5)$ & $-0.05(1)$  \\
  $0.7$ & $0.1064$& $0.75(8)$ & $-0.37(6)$ & $0.05(9)$ & $-0.05(6)$  \\
  $0.8$ & $0.1070$ & $0.76(3)$ & $-0.38(9)$ & $0.06(9) $ & $-0.06(1)$  \\
  $0.9$ & $0.1074$& $0.76(6)$ & $-0.39(7)$ & $0.07(5)$ & $-0.06(4)$  \\
  $1$ &$0.1075$ & $0.76(7)$ & $-0.39(9)$ & $0.07(7)$ & $-0.06(5)$  \\
  \hline
\end{tabular}
\end{center}
As an example, Fig. \ref{fig-fit} shows the fit (\ref{fit}) for
$B=0.5$ together with the corresponding values of $n
(f(0,n)-\tilde{f}(\infty))$ for integer values of $n$ in the
interval $[2,50]$. The fit (\ref{fit}) is in fact extremely good
for all values of $B$. More precisely, we have checked that the
values of $n (f(0,n)-\tilde{f}(\infty))-p_{\text{fit}}(n)$ for
integer $n$ are always smaller than $10^{-6}$ for all values of
$B$ included in the previous table.
\begin{figure}
\begin{center}
 \includegraphics[width=12cm,height=8cm,angle=0]{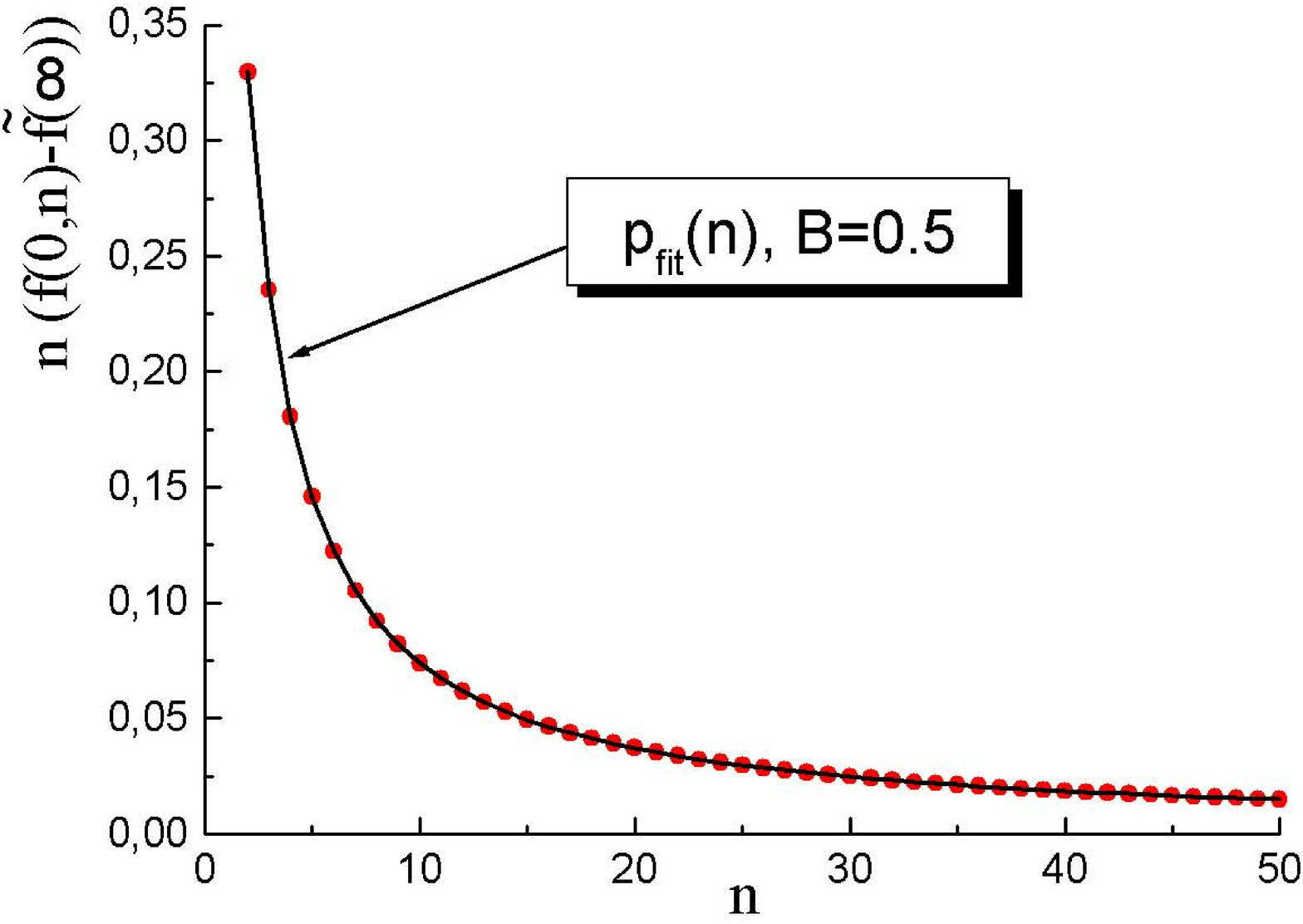}
\end{center}
    \caption{[Color online] For $B=0.5$, $p_{\text{fit}}(n)$ (\ref{fit}), and $n(f(0,n)-\tilde{f}(\infty))$ at integer values of $n$, in the interval $[2,50]$.}
\label{fig-fit}
\end{figure}
We are now in the position to compute $\tilde{f}(1)$ numerically.
Its value for a certain $B$ corresponds simply to adding up all
numbers in the corresponding row of the table above. The outcome
of this computation is:
\begin{center}
\begin{tabular}{|c|c|c|c|c|c|c|c|c|c|c|c|}
  \hline
$B$ & $0.02$ & $0.1$ & $0.2$ & $0.3$ & $0.4$ & $0.5$ & $0.6$ & $0.7$ & $0.8$& $0.9$&$1$ \\
  \hline
   $\tilde{f}(1)$ & $ 0.5(1)$ & $0.5(1)$  & $0.5(0)$ & $0.5(0)$ &$0.5(0)$  & $0.5(0)$ & $0.4(9)$ &$0.4(9)$  &$0.4(9)$  &$0.4(9)$ & $0.4(9)$ \\
  \hline
\end{tabular}
\end{center}
As expected,  all values obtained agree with the predicted value of $1/2$ within our numerical precision. Hence, the entanglement entropy can be computed to
\begin{equation}
   S_A(rm) = -\frac{1}{3}\log(\epsilon m) + U_{{\rm sinh-Gordon}}-\frc18 K_0(2rm) + O\lt(e^{-4rm}\rt)~.
\end{equation}

We do not know how to evaluate the constant $U$ yet, but the
methods developed by S.~Lukyanov \cite{finiteT} may be helpful.
\section{Generalization to theories with several particles and bound states}\label{manypar}

\noindent So far we have been dealing with theories with a single
particle spectrum and no bound states. In this section we wish to
investigate how our results for the entropy can be generalized to
situations in which the spectrum of the initial integrable QFT
consists of $\ell>1$ particles and bound states are still absent.
As before we will consider $n$ copies of the theory and therefore
it will be natural to label particles by two indices
\begin{equation}
    (\alpha,i) \quad \text{with} \quad \alpha=1,\ldots,\ell \quad \text{and}
    \quad i=1,\ldots,n\,,
\end{equation}
with the identification $(\alpha,i)\equiv (\alpha,i+n)$ for all
$i=1,\ldots,n$ and $\alpha=1,\ldots,\ell$. Denoting by $\Psi_{(
\alpha,i)}$ some fundamental field of the theory related to particle
$(\alpha,i)$, the exchange relations with the fields $\mathcal{T}$
and $\tilde{\mathcal{T}}$ can be written similarly as before:
\begin{eqnarray}
  \Psi_{(\alpha,i)}(y) \mathcal{T}(x) &=& \mathcal{T}(x)\Psi_{(\alpha,i+1)}(y)  \qquad x^{1}> y^{1}, \\
 \Psi_{( \alpha,i)}(y) \mathcal{T}(x) &=& \mathcal{T}(x)\Psi_{( \alpha,i)}(y)  \qquad x^{1}< y^{1}, \\
 \Psi_{( \alpha,i)}(y) \tilde{\mathcal{T}}(x) &=& \tilde{\mathcal{T}}(x)\Psi_{(\alpha,i-1)}(y)  \qquad x^{1}>
 y^{1},\\
 \Psi_{( \alpha,i)}(y) \tilde{\mathcal{T}}(x) &=& \tilde{\mathcal{T}}(x)\Psi_{(\alpha,i)}(y)  \qquad
 x^{1}< y^{1}.
\end{eqnarray}
Denoting by $S_{\alpha \beta}(\theta)$ with $ \alpha,\beta=1,\ldots,\ell$ the two-particle $S$-matrix of the original theory, the $S$-matrix of the $n$-sheeted theory can be written as:
\begin{eqnarray}
  S_{(\alpha,i) (\beta,i)}(\theta) &=& S_{\alpha\beta}(\theta)\quad \forall \quad \alpha, \beta , i,\\
 S_{(\alpha,i) (\beta,j)}(\theta) &=& 1 \quad\qquad \forall \quad \alpha,\beta,i,j \quad \text{with}
  \quad i \neq j\,.
\end{eqnarray}
The form factors axioms (\ref{kre}) for the operator $\tw$ still hold, with $\mu$ the double index $(\alpha,i)$, $\h\mu = (\alpha,i+1)$ and $\b\mu = (\b\alpha,i)$. If particles $\alpha,\beta$ in the original theory fuse to produce a bound state $\gamma$, then the $S$-matrix has a pole on the imaginary line of the physical sheet, say $iu_{\alpha\beta}^\gamma$ with $u_{\alpha\beta}^\gamma \in (0,\pi)$. Correspondingly, the form factors will possess extra poles with the requirements
\beqa &&
    -i\lim_{\varep\to0} \varep F_{n+1}^{\tw|(\alpha,i) (\beta,i) \mu_1\ldots\mu_{n-1}}
        (\theta+\frc{iu_{\alpha\beta}^\gamma}2-\varep,\theta-\frc{iu_{\alpha\beta}^\gamma}2+\varep,\theta_1,\ldots,\theta_{n-1}) \n && \qquad\qquad
        = \Gamma_{\alpha\beta}^\gamma F_{n+1}^{\tw|(\gamma,i)\mu_1\ldots\mu_{n-1}}(\theta_1,\ldots,\theta_{n-1})
\eeqa
where the so-called three-point coupling is
\beq
    \lt(\Gamma_{\alpha\beta}^\gamma\rt)^2 = -i\lim_{\theta\to iu_{\alpha\beta}^\gamma} (\theta-iu_{\alpha\beta}^\gamma) S_{\alpha\beta}(\theta)~.
\eeq

As usual, for finding solutions to these equations we must first construct minimal solutions of the two-particle form factor equations. They
satisfy the equations
\begin{eqnarray}
F_{\text{min}}^{\mathcal{T}|(\alpha,j) (\beta,k) }(\theta,
n)=F_{\text{min}}^{\mathcal{T}|(\beta,k) (\alpha,j) }(-\theta,
n)S_{(\alpha,j)  (\beta,k) }(\theta)=
F_{\text{min}}^{\mathcal{T}|( \beta,k) ( \alpha,j+1)}(2\pi
i-\theta, n),  \label{minix}
\end{eqnarray}
for all values of $j,k,\alpha$ and $\beta$. From the equations above it follows:
\begin{eqnarray}
  F_{\text{min}}^{\mathcal{T}|(\alpha,i) (\beta,i+k)}(\theta,n) &=&
  F_{\text{min}}^{\mathcal{T}|(\alpha,j)  (\beta, j+k) \,}(\theta,n) \qquad
\forall \quad i,j,k,\alpha,\beta\label{usex}\\
F_{\text{min}}^{\mathcal{T}|(\alpha,1)(\beta,j)}(\theta,
n)&=&F_{\text{min}}^{\mathcal{T}|(\beta,1)(\alpha,1)}(2\pi(j-1)i-\theta,
n )\qquad \forall \quad \alpha,\beta,j\neq 1.\label{use2x}
\end{eqnarray}
So, as before computing the minimal form factors of particles in
the first sheet is sufficient to determine all minimal form
factors of the theory. This minimal form factor must have no poles in the
extended strip $\text{Im}(\theta) \in [0, 2\pi n]$ and satisfies a
similar type of equations as for the case with $n=1$,
\begin{equation}
  F_{\text{min}}^{\mathcal{T}|(\alpha,1)(\beta,1)}(\theta, n)=F_{\text{min}}^{\mathcal{T}|(\beta,1)(\alpha,1)}(-\theta, n)S_{\alpha\beta}(\theta)=
  F_{\text{min}}^{\mathcal{T}|(\beta,1)(\alpha,1)}(-\theta+ 2\pi n i, n). \label{relx}
\end{equation}
From arguments completely analogous to those developed for the one
particle case, provided that the $S$-matrix of the original theory
admits an integral representation of the form
\begin{equation}
    S_{\alpha \beta}(\theta)=\exp \left[\int_{0}^{\infty} \frac{dt }{t} g_{\alpha\beta}(t) \sinh\left(\frac{t \theta}{i
    \pi}\right)\right],
\end{equation}
where $g_{\alpha \beta}(\theta)$ is a function which depends of the theory
under consideration with the property $g_{\alpha\beta}(t) = g_{\beta\alpha}(t)$ (that is, parity invariance), the minimal form factor is given by
\begin{equation}
    F_{\text{min}}^{\mathcal{T}|\alpha\beta}(\theta,n)=\exp\left[\frac{1}{2}\int_{0}^{\infty} \frac{dt}{t \sinh(n t)} g_{\alpha\beta}(t)
    \cosh\left({t}\left(n+\frac{i\theta}{\pi}\right)
    \right)\right].
\end{equation}

We are now in the position to obtain the full two-particle form
factors by including appropriate poles. In contrast to the  single
particle spectrum case not all two-particle form factors will have
poles. The kinematic and bound-state residue equations ensure that
the only singularities occur for form factors of the type
$F_{2}^{\mathcal{T}|(\alpha,i)(\bar{\alpha},j)}(\theta,n) $ for
any values of $i, j, \alpha$, or of the type
$F_{2}^{\mathcal{T}|(\alpha,i)(\beta,j)}(\theta,n) $ for any value
of $\alpha,\beta$ for which there is a non-zero three-point
coupling $\Gamma_{\alpha\beta}^\gamma$. If
$\Gamma_{\alpha\b\alpha}^\gamma=0$, then the pole structure of
these form factors is analogous to the one found for the one particle
case
\begin{equation}
    F_{2}^{\mathcal{T}|(\alpha,1) (\b\alpha,1)}(\theta)=\frac{
        \langle \mathcal{T}\rangle \sin\left(\frac{\pi}{n}\right)}{2 n \sinh\left(\frac{ i\pi - \theta}{2n}\right)\sinh\left(\frac{i\pi+\theta}{2n}\right)}
    \frc{F_{\text{min}}^{\mathcal{T}|(\alpha,1) (\b\alpha,1)}(\theta, n)}{F_{\text{min}}^{\mathcal{T}|(\alpha,1) (\b\alpha,1)}(i\pi, n)}.\label{fullx}
\end{equation}
In the presence of bound states, more factors need to be multiplied in order to account for the bound-state poles:
\begin{equation}\label{fullxbs}
    F_{2}^{\mathcal{T}|(\alpha,1) (\beta,1)}(\theta)=
        \frac{\lt(A + B \cosh\lt(\frc{\theta}n\rt) + C \cosh\lt(\frc{2\theta}n\rt) \rt)F_{\text{min}}^{\mathcal{T}|(\alpha,1) (\beta,1)}(\theta, n)
        }{\lt(\sinh\left(\frac{ i\pi - \theta}{2n}\right)\sinh\left(\frac{i\pi+\theta}{2n}\right)\rt)^{\delta_{\alpha,\b\beta}}
        \sinh\left(\frac{ iu_{\alpha\beta}^\gamma - \theta}{2n}\right)\sinh\left(\frac{iu_{\alpha\beta}^\gamma+\theta}{2n}\right)
    }
\end{equation}
where $A,B,C$ are constants that depend on $\alpha,\beta, n$. For all other form factors we have
\begin{equation}
F_{2}^{\mathcal{T}|(\alpha,1) (\beta,1)}(\theta)\propto
F_{\text{min}}^{\mathcal{T}|(\alpha,1) (\beta,1)}(\theta),\quad
\text{for}\quad \alpha\neq \bar{\beta}~,\quad \Gamma_{\alpha\beta}^\gamma =0.
\end{equation}

\subsection{Computation of the entropy}\indent \\

\noindent We can now proceed to the computation of the entropy along the lines described in the previous section. The two-point function in the two-particle approximation is
\beq\label{pss}
    \langle \mathcal{T}(r) \tilde{\mathcal{T}}(0)\rangle =
    \langle\mathcal{T}\rangle^2 \lt(1 + \frc{n}{8\pi ^2} \sum_{\alpha,\beta=1}^\ell
    \int\limits_{-\infty}^{\infty } \int\limits_{-\infty }^{\infty }d\theta_{1}d\theta _{2} f_{\alpha,\beta}(\theta_{12},n)\,
    e^{-r(m_\alpha\cosh\theta_1 + m_\beta\cosh \theta_2)} \rt)
\eeq
where $m_\alpha$ is the mass associated to particle type $\alpha$. The new function $f_{\alpha,\beta}(\theta,n)$ is given by
\begin{eqnarray}
 {\langle\mathcal{T}\rangle^{2}}f_{\alpha,\beta}(\theta,n) &=&
 \sum_{i=1}^{n}\left|F_{2}^{\mathcal{T} |(\alpha,1)(\beta,i)}(\theta,n)\right|^2 \\
   &=& \left|F_{2}^{\mathcal{T} |(\alpha,1)(\beta,1)}(\theta,n) \right|^2 +
  \sum_{j=1}^{n-1} \left|F_{2}^{\mathcal{T} |(\alpha,1)(\beta,1)}(-\theta+2\pi ij,n)\right|^2.
  \nonumber
\end{eqnarray}
For general integrable QFTs it is also possible that the twist
fields admit non-vanishing form factors both for odd and even
particle numbers. If that would be the case there would be an
extra contribution to the two-point function (\ref{pss}) coming
from the 1-particle form factor. However this would have no contribution to the entropy, since the 1-particle form
factors have no singularities in $\theta$ and therefore the
1-particle form factor contribution will exactly vanish at $n=1$.

For the computation of the entropy we have exploited the pole
structure of the form factors. In fact, from our analysis it
follows that only form factors containing poles will contribute to
the final value of $\tilde{f}_{\alpha,\beta}(1)$, where $\t{f}_{\alpha,\beta}(n)$ is the natural
analytic continuation of $f_{\alpha,\beta}(0,n)$ from $n=2,3,\ldots$ to $n\in[1,\infty)$.
More precisely, our analysis showed that only when poles collide do we have a non-zero contribution to the analytically-continued sum above at $n=1$ (recall that the first term vanishes at $n=1$). Since $0<u_{\alpha\beta}^\gamma<\pi$, only kinematic poles can collide at $n=1$, and this only occurs when $\theta=0$. Selecting out those terms that contain kinematic poles, we obtain the constraint that $\beta=\b\alpha$ in (\ref{pss}) for $n\sim1$, which allows us to perform one of the integrals as before:
\beq
    \langle \mathcal{T}(r) \tilde{\mathcal{T}}(0)\rangle \approx \langle\mathcal{T}\rangle^2\left(1 + \frac{n}{4\pi^2} \sum_{\alpha=1}^\ell
    \int\limits_{-\infty }^{\infty } d\theta \t{f}_{\alpha,\b\alpha}(\theta,n) K_{0}(2rm\cosh(\theta/2))\right)
    \qquad (n\sim1)
\end{equation}
for $\t{f}_{\alpha,\beta}(\theta,n)$ the natural analytic continuation of $f_{\alpha,\beta}(\theta,n)$. From here, the results of the previous section easily generalise to
\begin{equation}
    \tilde{f}_{\alpha,\b\alpha}(1)=\frac{1}{2}~,\quad \lt(\frc{\p}{\p n} \t{f}_{\alpha,\b\alpha}(\theta,n) \rt)_{n=1} = \frc{\pi^2}2 \delta(\theta)
\end{equation}
and
\begin{equation}
    S_A(rm)=-\frac{c}{3}\log(\epsilon m) + U -  \frc18 \sum_{\alpha=1}^\ell K_0(2rm_\alpha) + O\lt(e^{-3rm_1}\rt)
\end{equation}
where $m_1$ is the smallest mass and we admitted the possibility of non-zero three-particle form factors. In particular, the leading exponential correction is given by the mass of the lightest particles, and the coefficient is proportional to the number of such lightest particles in the spectrum.

\section{Conclusion} \label{conclusion}

 In this paper we have developed a program which, taking full advantage of
    quantum integrable model techniques, allows for the computation of the entanglement entropy of
    a connected region of a quantum one-dimensional integrable system with respect to the remaining part of the system.
    The starting point of
    our computation is the well-known ``replica-trick". Our program
    relies in the realization that starting with a quantum
    integrable model and constructing a new model consisting of
    $n$ non-interacting copies of the original theory, a new local
    quantum field theory is obtained which naturally possesses
    $\mathbb{Z}_n$ symmetry. Associated to this symmetry two twist
    fields $\mathcal{T}$ and $\tilde{\mathcal{T}}$ exist, whose two-point function is directly related
    to the entropy of the system. Through the ``replica-trick"
    mentioned above, the entropy is the derivative with respect to
    $n$ of the two-point function of the twist fields, evaluated at
    $n=1$.

Since the fields $\mathcal{T}$ and $\tilde{\mathcal{T}}$ are local
fields of the $n$-copy theory, their two-point function can be
computed by exploiting the form factor program for integrable
models. More precisely, it can be expressed as a sum for different
particle numbers over products of the form factors of the two
fields involved. This gives a large-distance expansion: computing the two-point function in the
two-particle approximation gives the leading behaviour at large distances.
This expansion is in fact expected to converge rapidly, and this
leading behaviour is often enough to describe the two-point
function up to relatively small distances.

The behaviour of the entropy as a function of the distance was
already well-known for very short and very large distances. For
$r\ll m^{-1}$ ($m$ being the mass of the lightest particle) that
behaviour is determined by the underlying CFT which describes the
integrable QFT in the ultraviolet limit. Thus, the entropy can be
computed explicitly by using CFT techniques
\cite{Calabrese:2004eu,Calabrese:2005in}. At large separations
$r\gg m^{-1}$ (in the infrared limit) the entropy is known to
saturate to a constant value. The main result of our work has been
to evaluate the first correction to the entropy in the infrared
limit, providing therefore a description of the behaviour of the
entropy in the intermediate region of values of $rm$. This
correction is obtained from the two-particle contributions to the
form factor expansion. In fact, the two-particle approximation
provides both the saturation value of the entropy, coming from the
disconnected part of the correlation function (the square of the
vacuum expectation value of the twist fields), and the exact first
corrections up to $O(e^{-3rm})$, coming from the two-particle
contributions. We also computed the exact value of the saturation in
the Ising model and showed it to be in good agreement with previous
numerical results \cite{Latorre2}.

The most surprising result of our analysis has been to establish
that the leading correction to the entropy at large $rm$ is in
fact a universal quantity, that is, it does not depend on the
particular scattering matrix of the model we started with, but
only on the spectrum of masses of the particles of the original
theory. It is quite remarkable that the entropy should encode so
explicitly crucial information about the theory both in its UV
regime (the central charge) and in the IR regime (the number of
light particles).

We have deduced this result from general arguments and checked it
explicitly for the Ising and sinh-Gordon models. The mathematical
reason for this ``universal behaviour" is clear, as the result is
directly related to the presence of kinematic poles in the
two-particle form factors. Only form factors having such poles do
contribute to the final result for the entropy, and their
individual contributions turn out to be theory-independent. The
presence of bound state poles does not change our conclusions.
However, we do not yet have a physical understanding of this
result.

It would be interesting to compute higher order corrections to the
entropy coming from states with higher particle numbers. In addition, an
interesting application of our main result emerges, namely given a
lattice model whose underlying QFT is unknown, a computation of
the entropy could reveal the number of light particles of the theory by
extracting the coefficient of the first exponential correction at
large distances. Another interesting route is to compute the entanglement
entropy of a disconnected region. This would involve higher-point
correlation functions of the twist fields introduced here, and would
be useful in verifying certain general properties of the entanglement entropy
which follow from its interpretation as the number of ``links'' between
the regions considered.

An additional result of our work has been to develop the form
factor program for branch-point twist fields. As a consequence of the
particular exchange relations between these fields and the
fundamental fields of the theory, the form factor consistency
equations for branch-point twist fields are different from those associated to
standard local fields. In particular, the crossing and kinematic
residue equations are modified. Since here we have only
been concerned with the two-particle form factors of branch-point twist fields,
an interesting open problem remains, namely to find closed
solutions to the form factor equations for arbitrary or at least
higher particle numbers. From the form factor equations we expect
these solutions to be given in terms of elementary symmetric
polynomials of the variables $e^{\theta_i /n}$. Also, it would be
interesting to extend our analysis to non-diagonal theories, such
as the sine-Gordon model.

In the Ising model, it is possible, for even $n$, to have an
explicit representation of the branch-point twist fields using
their relation to $U(1)$ twist fields in the free Dirac fermion
model as shown in appendix \ref{appising}. This provides a way to
explicitly evaluate Ising form factors, where many-particle form
factors are obtained from two-particle form factors by the usual
Wick's theorem. It also gives explicit representations of
two-point functions in terms of the known Painlev\'e transcendant
representations of the two-point functions of $U(1)$ twist fields.
It would be interesting to study the Ising and Dirac models
further from this viewpoint.

Finally, the notion of twist field as introduced in this work is
new in the context of integrable QFT and can still be generalized
further. It would be very interesting, for instance, to study
twist fields in integrable models which possess non-abelian
internal symmetries (such as $su(2)$). Developing the form factor
program for such fields is a possible future line of research.

\paragraph{Acknowledgments:}

B.D.~acknowledges support from an EPSRC post-doctoral fellowship,
grant number GR/S91086/01. J.C. ~acknowledges partial support from
EPSRC grant EP/D050952/1.

\appendix
\section{Form factors of branch-point twist fields from angular quantisation in the sinh-Gordon model}
\label{appshG}

It is possible to describe massive integrable quantum field theory
using angular quantisation \cite{freefield1,freefield2}.  Angular
quantisation is obtained when the model is quantized on radial
half-lines emanating from some fixed point in space. That is, the
Hilbert space is some subspace of the space of field
configurations on those radial half-lines, and time evolution is a
rotation around that point. Let us consider the sinh-Gordon model,
whose scattering matrix is given by (\ref{smatrix}). The angular
quantisation of the model is obtained from free bosonic modes
satisfying the canonical commutation relations \beq
    [\la_\nu,\la_{\nu'}] = f(\nu) \delta(\nu+\nu')
\eeq
with
\beq
    f(\nu) = \frc{2\sinh\frc{\pi B \nu}{4} \sinh\frc{\pi (2-B) \nu}{4}}{\nu\cosh\frc{\pi \nu}2}
\eeq
along with the zero modes
\beq
    [P,q]=-i~.
\eeq
The hamiltonian of the system, generating rotation around the origin, is written in terms of the modes as
\beq
    K = \int_0^\infty d\nu \frc{\nu}{f(\nu)} \la_{-\nu}\la_\nu~.
\eeq
One can easily calculate
\beq
    [K,\la_\nu] = -\nu\la_\nu~.
\eeq The angular Hilbert space, a Fock space over these modes,
will be denoted by ${\cal F} = \oplus_p {\cal F}_p$. It is naturally
decomposed into subspaces ${\cal F}_p$ with fixed eigenvalue $p$
of the operator $P$ (this determines the way the fundamental field
behave when  it approaches the origin 0, which is, in angular
quantisation, the far negative infinity of space).

The main advantage of this construction is that in some integrable
models,  an explicit embedding is known of the states in the usual
quantisation scheme (with Hilbert space ${\cal H}$) into the space
of operators acting on ${\cal F}$. This embedding goes as follows
\cite{freefield2}. One first identifies vacuum expectation values
in ${\cal H}$ with traces on ${\cal F}$, as a simple consequence
of the change of quantisation scheme: \beq\label{trace}
    \bra \vac| A |\vac\ket = \frc{\Tr\lt(e^{-2\pi K} \iota(A) \rt)}{\Tr\lt(e^{-2\pi K} \rt)} \equiv \bra\bra \iota(A) \ket\ket
\eeq where on the left-hand sides $A$ stands for an operator that
is  the representation on ${\cal H}$ of products of local fields,
and on the right-hand side $\iota(A)$ stands for the
representation on the angular-quantisation Hilbert space ${\cal
F}$ of the same fields. Then, there are operators $Z(\theta)$
acting on ${\cal F}$ such that products of them correspond to
asymptotic states of ${\cal H}$: \beq
    \bra \vac| A |\theta_1,\ldots,\theta_k\ket = \bra\bra \iota(A) Z(\theta_1) \cdots Z(\theta_k)\ket\ket~.
\eeq
The operators $Z(\theta)$ are defined as follows. We have first
\beq\label{Lambda}
    \Lambda^\eta(\theta) = :e^{-i\eta\int d\nu \la_\nu e^{i\nu(\theta-i\pi/2)}}:
\eeq
for $\eta=\pm$, where the normal-ordering is with respect to the trace in (\ref{trace}), and we define
\beq
    Z(\theta) = -iC\lt[ e^{\frc{i\pi P}Q} \Lambda^+(\theta+i\pi/2)-e^{-\frc{i\pi P}Q} \Lambda^-(\theta-i\pi/2)\rt]
\eeq
with
\beq\label{C}
    C = \frc1{\sqrt{\sin\frc{\pi B}2}}\exp\lt[\int_0^\infty\frc{dt}t\frc{\sinh\frc{Bt}{4}\sinh\frc{(2-B)t}{4}}{\sinh t\cosh\frc{t}2}\rt]~.
\eeq The representation $\iota(A)$ on ${\cal F}$ of any field at
the origin that is local with respect to the fundamental field
commutes with these operators; this constraint is expected to be
sufficient to fix the set of all such fields in angular
quantisation (which excludes twist fields - they are discussed
below).

For instance, form factors of exponential fields $e^{a\varphi}$ at
the origin are obtained by choosing $\iota(A)$ to  be a projector
on $P=a$, times the vacuum expectation value $\bra
e^{a\varphi}\ket$. In particular, the identity operator is the
projector on $P=0$. Evaluating the traces is simple using \beq
    \bra\bra \la_\nu \la_{\nu'}\ket\ket = \frc{f(\nu)}{1-e^{-2\pi \nu}} \delta(\nu+\nu')
\eeq obtained from cyclic properties of the trace. From the usual
formula for evaluating averages  of normal-ordered exponentials of
free modes, \beq
    \bra\bra :e^{\int d\nu \lambda_\nu a(\nu)}:\ :e^{\int d\nu \lambda_\nu b(\nu)}: \ket\ket =
    e^{\int d\nu d\nu' a(\nu) b(\nu') \bra\bra \lambda_\nu \lambda_{\nu'}\ket\ket}~,
\eeq
we obtain
\beqa &&
    \bra\bra \Lambda^{\eta_1} \lt(\theta_1 + \eta_1 \frc{i\pi}2\rt) \Lambda^{\eta_2} \lt(\theta_2 + \eta_2 \frc{i\pi}2\rt) \ket\ket
    = \n && \quad
    \exp\lt[-2\eta_1\eta_2 \int_0^\infty \frc{dt}t \frc{\sinh \frc{B t}{4} \sinh \frc{(2-B)t}{4}}{\sinh t \cosh  \frc{t}2} \cosh t\lt(1+i\frc{\theta_1-\theta_2}\pi
    -\frc12(\eta_1-\eta_2)\rt)\rt]~.
\eeqa
Two-particle form factors of exponential fields are then given by
\beq
    \bra \vac| e^{a\varphi} |\theta_1,\theta_2\ket = \bra e^{a\varphi}\ket C^2 \sum_{\eta_{1,2} = \pm} e^{\lt(\frc{i\pi a}{Q}-\frc{i\pi}2\rt)(\eta_1+\eta_2)}
    \bra\bra \Lambda^{\eta_1} \lt(\theta_1 + \eta_1 \frc{i\pi}2\rt) \Lambda^{\eta_2} \lt(\theta_2 + \eta_2 \frc{i\pi}2\rt) \ket\ket
\eeq
and the constant $C$ ensures that the value of the kinematic residue is correct.

Angular quantisation is also useful for studying twist fields.
Since the ``space'' (or equal-time slices) of angular quantisation
is just the half line, a twist field associated to a symmetry
action $\sigma$ is just the operator for the symmetry action
itself on the angular Hilbert space: \beq\label{twangular}
    \iota(\tw_\sym(0)) = \bra\tw_\sym\ket \sym_{{\cal F}}
\eeq
and this also projects on the subspace with $P=0$. Note that the symmetry action applied on a field, $[\sym_{{\cal F}},\iota(\Or(x\neq0))]$, inside the trace on the angular Hilbert space, does produce the symmetry transformation of the field $\Or(x)$: since there is a commutator, there are two branches inserted at slightly different angular times, and the two branches cancel each other except around the point $x$. Also, these two branches can be deformed into one branch from $x=-\infty$ to $x=\infty$, so this is really the same object as the operator for the symmetry action on ${\cal H}$. When there is no commutator, that is when just $\sym_{{\cal F}}$ is inserted inside the trace, then this does not produce symmetry transformation, rather it corresponds to the insertion of a local field, the twist field.

We now consider the $n$-copy sinh-Gordon model: the new angular
quantisation Hilbert space is just $\oplus_{j=1}^n {\cal
F}^{(n)}$, where all ${\cal F}^{(n)}$ are isomorphic to ${\cal
F}$, and the asymptotic state operators are just $Z_j(\theta)$,
for $j=1,2,\ldots,n$. They are built as before out of bosonic
modes $\la_{j,\nu},\; j=1,2,\ldots,n$ that commute for different
values of $j$. The local twist field $\tw=\tw_\sym$ associated to
the symmetry $\sym: j\leftrightarrow j+1\ {\rm mod} \ n$ is then
just given by (\ref{twangular}).

Form factors of $\tw$ are then
\beq\label{fftw}
    \bra\vac| \tw|\theta_1,\ldots,\theta_k\ket_{\mu_1,\ldots,\mu_k} =
    \bra\tw\ket \t{C}^k \bra\bra Z_{\mu_1}(\theta_1) \cdots Z_{\mu_k}(\theta_k) \ket\ket_\sym
\eeq
with
\beq
    \bra\bra \cdots \ket\ket_\sym = \frc{\bra\bra \sym_{{\cal F}} \cdots \ket\ket}{\bra\bra \sym_{{\cal F}}\ket\ket}~.
\eeq
We now take the operators $Z_{\mu}(\theta)$ to be composed of exponential operators normal-ordered with respect to this new trace, in order for the computation of the trace of their products to go as before. Such a change of normal-ordering, for exponential of free modes, just changes the normalisation, and the operators $\Lambda_\mu^\pm(\theta)$ all get the same normalisation change. The constant $\t{C}$ has been introduced in order to account for this, and can be fixed by the requirement that the correct value of the kinematic residue is obtained from the two-particle form factors evaluated below (it could also be determined directly by a Bogoliubov transformation).

Using the fact that the symmetry acts like $\sym_{{\cal F}} \la_{j,\nu} \sym_{{\cal F}}^{-1} = \la_{j+1,\nu}$ (with $j+n \equiv j$) and cyclic properties of the trace, we can evaluate the following averages:
\beq
    \bra\bra \la_{j,\nu} \la_{1,\nu'}\ket\ket_\sym = \frc{e^{-2(j-1)\pi \nu} f(\nu)}{1-e^{-2n\pi \nu}} \delta(\nu+\nu')\qquad(j=1,2,\ldots,n)
\eeq
Note that with $j=1$, the average is the one obtained by changing the angle around the origin to $2\pi n$. This is naturally expected: the branch points represented by these branch-point twist fields are just negative-curvature conical singularities with an angle of $2\pi n$. For $j>1$, the formula says that in evaluating form factors with two particles belonging to different copies, one can replace $\la_{j,\nu}$ by $e^{-2(j-1)\pi \nu}\la_{1,\nu}$. From formula (\ref{Lambda}), it is clear that this is equivalent to changing the particle $l$, of type $j_l$, to type $1$, and shifting its rapidity $\theta_l$ to $\theta_l+2\pi i (j_l-1)$, as long as the resulting integral representation for the form factor is convergent after this shift. This is so if the shift keeps the rapidity in the strip ${\rm Im}(\theta_1-\theta_2)\in[0,2\pi n]$, so that this shift is allowed if it is the first particle, $l=1$ with rapidity $\theta_1$, that is changed. Hence, we find
\beq
    F_2^{\tw|j1}(\theta_1-\theta_2) = F_2^{\tw|11}(\theta_1-\theta_2+2\pi i(j-1))
\eeq
which is in agreement with (\ref{use2}) (this equation holds for the whole form factor, not just the minimal form factor).

Hence it is sufficient to evaluate form factors involving particles on the same copy. The evaluation of traces involved in calculating these form factors goes as above, and we find
\beqa &&
    \bra\bra \Lambda^{\eta_1}_1 \lt(\theta_1 + \eta_1 \frc{i\pi}2\rt) \Lambda^{\eta_2}_1 \lt(\theta_2 + \eta_2 \frc{i\pi}2\rt) \ket\ket_\sym
    = \n && \quad
    \exp\lt[-2\eta_1\eta_2 \int_0^\infty \frc{dt}t \frc{\sinh \frc{B t}{4} \sinh \frc{(2-B)t}{4}}{\sinh nt \cosh \frc{t}2} \cosh t\lt(n+i\frc{\theta_1-\theta_2}\pi
    -\frc12(\eta_1-\eta_2)\rt)\rt]~.
\eeqa
which gives
\beqa
    \bra \vac| \tw |\theta_1,\theta_2\ket_{1,1} &=& \bra\tw\ket (\t{C}C)^2 \sum_{\eta_{1,2} = \pm} e^{-\frc{i\pi}2(\eta_1+\eta_2)}
    \bra\bra \Lambda^{\eta_1}_1 \lt(\theta_1 + \eta_1 \frc{i\pi}2\rt) \Lambda^{\eta_2}_1 \lt(\theta_2 + \eta_2 \frc{i\pi}2\rt) \ket\ket_\sym \n
    &=& \bra\tw\ket
    (\t{C}C)^2 \lt(-2F_{\text{min}}^{\tw|11}(\theta) + \frc1{F_{\text{min}}^{\tw|11}(\theta+i\pi)} + \frc1{F_{\text{min}}^{\tw|11}(\theta-i\pi)}\rt)
\eeqa
(with $\theta=\theta_1-\theta_2$) where $F_{\text{min}}^{\tw|11}(\theta)$ is given by (\ref{int}). This can be seen to reproduce (\ref{full}) if we choose
\beq
    (\t{C}C)^2 =  \frc{-iC \sqrt{\sin\frc{\pi B}2}}{2\lt(\cos\frc{\pi B}4 + \sin\frc{\pi B}4 - 1\rt)}~,
\eeq
using the properties
\beq
    \frc1{F_{\text{min}}^{\tw|11}(\theta+i\pi)} = f(\theta) F_{\text{min}}^{\tw|11}(\theta)~,\quad \frc1{F_{\text{min}}^{\tw|11}(\theta-i\pi)}
    = f(2\pi in-\theta) F_{\text{min}}^{\tw|11}(\theta)
\eeq
with
\beq
    f(\theta) = \frc{\sin\frc{\pi}{2n} \lt(\frc{B}2 + 1 + \omega\rt) \sin\frc\pi{2n}\lt(-\frc{B}2 + 2+\omega\rt)}{
    \sin\frc{\pi}{2n} \lt(1 + \omega\rt) \sin\frc\pi{2n}\lt(2+\omega\rt)}
\eeq
and
\beq
    \omega = n+\frc{i\theta}\pi~.
\eeq

\section{Vacuum expectation values of branch-point twist fields in the Ising model}
\label{appising}

In this appendix, we use the relation between branch-point twist
fields in the $n$-copy Ising  model and the $n$ independent $U(1)$
twist fields in the $n$-copy free massive Dirac theory, in order
to deduce vacuum expectation values of the former from the known
formulas for those of the latter. Note that this derivation is
very similar to that employed in \cite{casini1}.

The symmetry $\sigma: i \mapsto i+1 \ {\rm mod} \ n$  associated
to the branch-point twist field in the $n$-copy Ising model can
certainly be diagonalised on the particle eigenstates. It would be
natural to associate the resulting diagonal elements with elements
of a $U(1)$ symmetry group, and to interpret the branch-point
twist field as a product of the corresponding $n$ independent
$U(1)$ twist fields. This, of course, is unnatural in the $n$-copy
Ising model, since there is no $U(1)$ symmetry in the individual
copies. The trick is to further {\em double} the $n$-copy Ising
model in order to make it into an $n$-copy Dirac theory. Denoting
the fundamental real Majorana fermion fields by $\psi_{a,j},\;
\psi_{b,j},\; \b{\psi}_{a,j},\; \b{\psi}_{b,j}$ for
$j=1,2,\ldots,n$ and fundamental Dirac spinor fermion field
$\Psi_j=\mato{c} \Psi_{R,j} \\ \Psi_{L,j} \matf$, we have the
relations \beq
    \Psi_{R,j} = \frc1{\sqrt2} (\psi_{a,j} + i\psi_{b,j})~, \quad \Psi_{L,j} = \frc1{\sqrt2} (\b\psi_{b,j} - i\b\psi_{a,j})~.
\eeq Each copy of the Dirac fermion has a $U(1)$ symmetry: it is
the symmetry under rotation between copies $a$ and $b$ of the
Ising model, which occurs because the Ising model is quadratic. In
the $n$-copy Dirac theory there is also an extra symmetry under
$SU(n)$ transformations of the multiplet $\mato{c} \Psi_1 \\
\cdots \\ \Psi_n\matf$, again because the theory is quadratic.
Diagonalising the branch-point twist field $\tw_{{\rm Dirac}}$ in
the Dirac theory, which is done with a $SU(n)$ transformation as
is seen below, it is possible to interpret the new basis as $n$
new independent Dirac fermions. It is then possible that the
branch-point twist field can be simply written as a product of $n$
independent $U(1)$ twist fields acting on these independent
fermions. This is useful for the Ising model, because by
uniqueness of the branch-point twist field (characterized by its
basic branch-point property and its dimension), we have (recall
that the central charge of the Dirac theory is 1) \beq
    \tw_{{\rm Dirac}} = \tw_{a} \otimes \tw_b
\eeq where $\tw_a$ and $\tw_b$ are the branch-point twist fields
in the copies $a$ and $b$ of the $n$-copy Ising model
respectively.

In order to go into the details, we need to be more precise about
the $SU(n)$ transformation that diagonalises $\tw_{{\rm Dirac}}$.
It would be natural to expect that the Fourier transform
$\Psi^{(q)} = \frc1n \sum_{j=1}^{n} e^{2\pi i j q} \Psi_j$ is the
appropriate transformation, but there is an important subtlety:
the various copies of the Dirac fermions {\em commute} amongst
each other, whereas to form new fermions with linear combinations
of them, they need to {\em anti-commute}. This is clear from our
choice of scattering matrix $S_{jk}(\theta) = 1$ amongst different
copies $j\neq k$. Of course, from the viewpoint of the Hilbert
space this is only a choice of basis, and it is possible to define
a new basis with scattering matrix $-1$ amongst different copies.
There are many ways of doing that, all with the same result. We
will choose the following (here with a two-particle example for
simplicity): \beq
    |\theta_1 \theta_2\ket_{j_1,j_2}^{{\rm ac}} = \lt\{ \ba{ll} |\theta_1 \theta_2\ket_{j_1,j_2} & j_1\leq j_2 \\ -|\theta_1 \theta_2\ket_{j_1,j_2} & j_1>j_2 \ea \rt.
\eeq Written using annihilation and creation operators for the
basis  with upper index ${\rm ac}$, fermions of different copies
anti-commute (and this is the only change). In the new basis, the
symmetry $\sigma$ acts as follows: \beq
    \sigma \Psi^{{\rm ac}}_j = \lt\{ \ba{ll} \Psi^{{\rm ac}}_{j+1} & j=1,\ldots,n-1 \\ -\Psi^{{\rm ac}}_{1} & j=n~. \ea \rt.
\eeq The extra minus sign is understood as follows: in a non-zero
correlation function,  there are an even number of fermion fields.
Let us order them in increasing copy label. There will be, say,
$k$ fields with copy label $n$ at the right. Applying the symmetry
makes them into copy $1$, and they are the only fields that break
the order of increasing copy label. Bringing them back to the
beginning gives a minus sign if they are of odd number. This is
cancelled by the extra minus sign above.

Writing $\sigma$ in matrix form, $\sum_{k=1}^n \sigma_{jk} \Psi_k
= \sigma \Psi_j$,  we find that the eigenvalues $\lambda$,
$\sum_{k=1}^n \sigma_{jk} v_k = \lambda v_j $, satisfy \beq
    \lambda^n = -1\Rightarrow \lambda = e^{\frc{i\pi p}n}~, \quad p\quad \mbox{odd}~.
\eeq
The explicit $SU(n)$ transformation is
\beq
    \Psi^{(p)} = \sum_{j=1}^n \lt(e^{-i \pi pj/n} - e^{-i \pi p(1+j/n)}\rt) \Psi_j~.
\eeq

Hence, we may expect that, for $n$ even,
\beq\label{idtw}
    \tw_{{\rm Dirac}} = \prod_{q=1}^{\frc{n}2} \Or_{\frc{2q-1}{2n}}^{(2q-1)} \prod_{q=1}^{\frc{n}2} \Or_{-\frc{2q-1}{2n}}^{(-2q+1)}
\eeq where $\Or_\alpha^{(p)}$ is the $U(1)$ twist field acting
non-trivially only on the fermion  fields $\Psi^{(p)}$, associated
with the $U(1)$ element $e^{2\pi i \alpha}$, for $\alpha\in[0,1]$.
The dimension of these twist fields is $\alpha^2$, and the choice
of the odd values of $p$ above forming the product is dictated by
the requirement of having the lowest total dimension. The
dimension of the product of fields on the right hand side of
(\ref{idtw}) is \beq
    2\sum_{q=1}^{\frc{n}2} \lt(\frc{2q-1}{2n}\rt)^2 = \frc{1}{12}\lt(n-\frc1n\rt)
\eeq which agrees with the CFT prediction (\ref{scdim}) for
central charge $c=1$.  By uniqueness of the branch-point twist
field, the factorisation above must be the correct one. We checked
in the case $n=2$ that our form factors agree with formula
(\ref{idtw}) along with the known form factors form $U(1)$-twist
fields \cite{SchroerT78}. For $n$ odd, similar calculations show
that there are many lowest-dimension factorized operators, and
that their dimension does not agree with the CFT prediction (it is
higher). Hence, we expect that in this case, no such factorisation
exists. This is not a problem, as we are interested in the
analytic expression for the vacuum expectation value, which can be
obtained from even values of $n$.

We can now read off the expectation values of $\Or_\alpha$ from
\cite{Zamounp,LukyanovZ}: \beq
    \bra \Or_\alpha\ket = \lt(\frc{m}2\rt)^{\alpha^2} \exp\int_0^\infty \frc{dt}t \lt( \frc{\sinh^2\alpha t}{\sinh^2 t} - \alpha^2 e^{-2t}\rt) =
     \lt(\frc{m}2\rt)^{\alpha^2} \frc1{G(1-\alpha)G(1+\alpha)}
\eeq
which gives (\ref{vevtw}) (here $G(z)$ is Barnes' G-function).

\section{The function $\t{f}(\theta,n)$}
\label{appf}

We now derive the full analytic continuation $\t{f}(\theta,n)$  in
an integral representation. For simplicity, we will assume that
$F_{2}^{\mathcal{T} |11}(\theta,n)$ is $0$ at $\theta=0$ and that
it vanishes exponentially as $|\theta|\to\infty$, as is observed
in the Ising and the sinh-Gordon models. Let us write \beq
    \t{f}(\theta,n) = \sum_{j=0}^{n-1} s(\theta,j)
\eeq
with $s(\theta,j)= F_{2}^{\mathcal{T} |11}(-\theta+2\pi i j,n) \lt(F_{2}^{\mathcal{T} |11}\rt)^*(-\theta-2\pi i j,n)$. Consider now the following contour integral:
\beq
\int_C\pi\cot(\pi z)s(\theta,z)\frac{dz}{2\pi i}
\eeq
where $C$ is the closed rectangular contour with vertices $(n-iL,n+iL,iL,-iL)$. Then, $s(\theta,z)$ decays exponentially as ${\rm Im}\,z\to\pm\infty$, and the contributions from the horizontal segments vanish as $L\to\infty$. Thanks to the quasi-periodicity of the integrand, $s(\theta,z+n) = S(\theta-2\pi iz) S(\theta+2\pi i z) s(\theta,z)$, the contributions from the vertical pieces amount to
\beq
    \int_{-i \infty}^{i \infty}(S(\theta-2\pi iz) S(\theta+2\pi i z)-1)\pi\cot(\pi z)s(z)\frac{dz}{2\pi i}~.
\eeq
The residues from the simple poles at $z=j$, $j=1,\ldots,n-1$ sum to $f(\theta,n)$. There are also simple poles of $s(z)$ at $z=\frac12\pm\frc{\theta}{2\pi i}$ and $z=n-\frac12\pm\frc{\theta}{2\pi i}$ -- the kinematic singularities. They evaluate to (for real $\theta$)
\beq
    \tanh\lt(\frc{\theta}2\rt)\ {\rm Im} \lt( F_{2}^{\mathcal{T} |11}(-2\theta+i\pi,n) - F_{2}^{\mathcal{T} |11}(-2\theta+2\pi i n - i\pi,n)\rt) ~.
\eeq
At $\theta=0$ this gives $-1/2$ using the kinematic residue equation, as it should. For any non-zero $\theta$, it vanishes as $n\to1$, like $(n-1)^2$. As an analytic function of $\theta$, there are simple poles at $\theta = \pm i\pi(n-1)$, which are those that give the main contribution as $n\to1$ for $\theta$ near to 0. These poles are
\[
    \frc{i\cot\lt(\frc{\pi n}2\rt) }{2(\theta+i\pi(n-1))}~,\quad - \frc{i\cot\lt(\frc{\pi n}2\rt) }{2(\theta-i\pi(n-1))}
\]
which indeed gives the behavior (\ref{tfnm}), with $\t{f}(1) = 1/2$.

The full analytic continuation can now be written:
\beqa
   \t{f}(\theta,n) &=& \tanh\lt(\frc{\theta}2\rt) \ {\rm Im} \lt( F_{2}^{\mathcal{T} |11}(-2\theta+i\pi,n) - F_{2}^{\mathcal{T} |11}(-2\theta+2\pi i n - i\pi,n)\rt) \\ &&
        - \frc1{4i\pi}\int_{-\infty}^{\infty}  \coth\lt(\frc\beta2\rt) (S(\theta-\beta)-S(\theta+\beta)) F_{2}^{\mathcal{T} |11}(-\theta+\beta,n)
        \lt(F_{2}^{\mathcal{T} |11}(\theta+\beta,n)\rt)^* d\beta ~. \no
\eeqa
In particular, at $\theta=0$, this specializes to
\beq\label{tfn}
    \t{f}(n) = \frc12 - \frc1{2\pi}\int_{-\infty}^{\infty} {\rm Im}(S(-\beta)) \coth\lt(\frc\beta2\rt) |F_{2}^{\mathcal{T} |11}(\beta,n)|^2 d\beta~.
\eeq

\section{Computation of  $\tilde{f}(\infty)$ for the sinh-Gordon model }
\label{finf}

In this appendix we present a detailed derivation of the equation
of the line that provides the large $n$ behaviour of the function
$f(0,n)$ (and by analytic continuation that of $\tilde{f}(n)$)
in the sinh-Gordon model. The computation can be carried out as
follows: the limits of the various factors entering the
two-particle form factor at $\theta=2 \pi i (j-1)$ are given by
\begin{eqnarray}
  \lim_{n \rightarrow \infty} F_{\text{min}}^{\mathcal{T}|11}(2\pi i (j-1), n) &=& (j-1) \frac{\Gamma\left(j-\frac{1}{2}-\frac{B}{4}\right)
  \Gamma\left(j-1+\frac{B}{4}\right)
  }{\Gamma\left(j-\frac{B}{4}\right)\Gamma\left(j-\frac{1}{2}+\frac{B}{4}\right)},\quad
  \text{for}\quad j\ll n,\nonumber \\
 \lim_{n \rightarrow \infty} F_{\text{min}}^{\mathcal{T}|11}(i \pi, n) &=& \frac{1}{2}\frac{\Gamma\left(1-\frac{B}{4}\right)
  \Gamma\left(\frac{1}{2}+\frac{B}{4}\right)
  }{\Gamma\left(\frac{3}{2}-\frac{B}{4}\right)\Gamma\left(1+\frac{B}{4}\right)},\nonumber\\
  \lim_{n \rightarrow \infty}\frac{\sin\left(\frac{\pi}{n}\right)}
  {2 n \sinh\left(\frac{i \pi (2j-1)}{2n} \right)\sinh\left(\frac{i \pi (3-2j)}{2n}
  \right)}&=& \frac{2}{\pi (2j-1)(2j-3)}, \quad
  \text{for}\quad j\ll n,
\end{eqnarray}
where for the first two functions, the limit can be easily
evaluated by employing the integral representation (\ref{int}) of
the minimal form factor. Putting all factors together we obtain
\begin{eqnarray}
 && \lim_{n \rightarrow \infty} F_{2}^{\mathcal{T} |11}(2\pi i(j-1),n)\nonumber\\
 && =\frac{4 (j-1)}{\pi (2j-1)(2j-3)}\frac{\Gamma\left(j-\frac{1}{2}-\frac{B}{4}\right)
  \Gamma\left(j-1+\frac{B}{4}\right)\Gamma\left(\frac{3}{2}-\frac{B}{4}\right)
  \Gamma\left(1+\frac{B}{4}\right)
  }{\Gamma\left(j-\frac{B}{4}\right)\Gamma\left(j-\frac{1}{2}+\frac{B}{4}\right)\Gamma\left(1-\frac{B}{4}\right)
  \Gamma\left(\frac{1}{2}+\frac{B}{4}\right)}, \label{ap}
\end{eqnarray}
for $j \ll n$. In order to compute $f(0,\infty)$ we still need to
perform the sum over $j$ of the function above. Since the sum goes
up to $j=n$ and in (\ref{ap}) we have assumed that $j \ll n$, it
is not guaranteed that (\ref{ap}) will provide a good
approximation for the whole range of values $j$. It turns out that
(\ref{ap}) is the right function to consider for two reasons:
first, computing $ F_{2}^{\mathcal{T} |11}(2\pi i(j-1),n)$
numerically for various values of $j$ and $n$ one quickly realizes
that it is only non-zero for values of $j$ around 2 or around $n$;
second, due to the periodicity of the form factor at $\theta=0$,
it turns out that the value of the function (\ref{ap}) does not
change when $j \rightarrow n-j +2$. Therefore, if we call the
function (\ref{ap}) $k(j)$, we can replace $\sum_{j=2}^{n} k(j)
\rightarrow 2\sum_{j=2}^{\infty} k(j)$ for $n \rightarrow \infty$.
We then obtain
\begin{equation}
  f(0,\infty)=  \tilde{f}(\infty
    )=\frac{32}{\pi^2}\left[\frac{\Gamma\left(\frac{3}{2}-\frac{B}{4}\right)\Gamma\left(1+\frac{B}{4}\right)}{\Gamma\left(1-\frac{B}{4}\right)
  \Gamma\left(\frac{1}{2}+\frac{B}{4}\right)
  }\right]^2
  A(B),\label{slope}
\end{equation}
with
\begin{equation}\label{a}
    A(B)=\sum_{j=2}^{\infty}\frac{(j-1)^2}{ (2j-1)^2(2j-3)^2}\left[\frac{\Gamma\left(j-\frac{1}{2}-\frac{B}{4}\right)
  \Gamma\left(j-1+\frac{B}{4}\right)
  }{\Gamma\left(j-\frac{B}{4}\right)\Gamma\left(j-\frac{1}{2}+\frac{B}{4}\right)}\right]^2.
\end{equation}
Employing the definition of the Pochhammer symbol
\begin{equation}
(a)_n = \frac{\Gamma(a+n)}{\Gamma(a)}=a (a+1)(a+2) \ldots (a+n-1),
\end{equation}
we can rewrite the sum above as
\begin{equation}\label{a2}
    A(B)=\left[\frac{\Gamma\left(\frac{3}{2}-\frac{B}{4}\right)
  \Gamma\left(1+\frac{B}{4}\right)
  }{\Gamma\left(2-\frac{B}{4}\right)\Gamma\left(\frac{3}{2}+\frac{B}{4}\right)}\right]^2 \sum_{j=2}^{\infty} \frac{(j-1)^2}{(2j-1)^2
  (2j-3)^2} \left[\frac{\left( \frac{3}{2}-\frac{B}{4}\right)_{j-2}\left( 1+\frac{B}{4}\right)_{j-2}}
  {\left( 2-\frac{B}{4}\right)_{j-2}\left(
  \frac{3}{2}+\frac{B}{4}\right)_{j-2}}\right]^2,
\end{equation}
and carry it out analytically. The result is given in terms of
generalized hypergeometric functions, which are defined as
\begin{equation}\label{fpq}
   _{p}F_{q}\left[ \begin{array}{c}
     a_1, a_2, \ldots, a_p ; z \\
     b_1, b_2, \ldots, b_q \\
   \end{array}\right]=\sum_{k=0}^{\infty} \frac{(a_1)_k (a_2)_k \ldots (a_p)_k}{(b_1)_k (b_2)_k \ldots
   (b_q)_k}\frac{z^k}{k!}.
\end{equation}
We find
\begin{eqnarray}
&& \sum_{j=2}^{\infty} \frac{(j-1)^2}{(2j-1)^2
  (2j-3)^2}  \left[\frac{\left( \frac{3}{2}-\frac{B}{4}\right)_{j-2}\left( 1+\frac{B}{4}\right)_{j-2}}
  {\left( 2-\frac{B}{4}\right)_{j-2}\left(
  \frac{3}{2}+\frac{B}{4}\right)_{j-2}}\right]^2 =\frac{1}{9}  \,_{7}F_{6}\left[ \begin{array}{c}
   \frac{1}{2}, \frac{1}{2}, 1, \alpha,\alpha,\beta,\beta; 1 \\
     \frac{5}{2},\frac{5}{2},\kappa,\kappa,\sigma,\sigma \\ \end{array}\right] \nonumber \\
     && + \frac{1}{75} \left(\frac{\alpha\beta}{\kappa\sigma}\right)^2\,_{7}F_{6}\left[ \begin{array}{c}
    \frac{3}{2}, \frac{3}{2}, 2, \alpha+1,\alpha+1,\beta+1,\beta+1; 1 \\
     \frac{7}{2},\frac{7}{2},\kappa+1,\kappa+1,\sigma+1,\sigma+1 \\
     \end{array}\right]\\
     && + \frac{2}{1225}\left(\frac{\alpha\beta}{\kappa\sigma}\right)^2\frac{(8+B)^2(B-10)^2}{(10+B)^2(B-12)^2}\,_{7}F_{6}\left[ \begin{array}{c}
    \frac{5}{2}, \frac{5}{2}, 3, \alpha+2,\alpha+2,\beta+2,\beta+2; 1 \\
     \frac{9}{2},\frac{9}{2},\kappa+2,\kappa+2,\sigma+2,\sigma+2 \\
     \end{array}\right],\nonumber
\end{eqnarray}
with
\begin{equation}
    \alpha=\frac{3}{2}-\frac{B}{4},\quad \beta=1+\frac{B}{4},\quad
    \kappa=2-\frac{B}{4},\quad \sigma=\frac{3}{2}+\frac{B}{4}.
\end{equation}

\bibliographystyle{phreport}
\small

\end{document}